\documentstyle[prd,eqsecnum,aps]{revtex}
\begin{document}
%\preprint{gr-qc/yymmddd}
\title{Head-on collisions of black holes: the particle limit}
\author{Carlos O. Lousto\thanks{Electronic Address:
lousto@mail.physics.utah.edu} and Richard H. Price}
\address{
Department of Physics, University of Utah, Salt Lake City, UT 84112}
\maketitle
\begin{abstract}
We compute gravitational radiation waveforms,
spectra and energies for a point particle of mass $m_0$
falling from rest at radius $r_0$ into a Schwarzschild hole
of mass $M$. This radiation is found to lowest order in $(m_0/M)$
with the use of a Laplace transform.
In contrast with numerical relativity results for head-on
collisions of equal-mass holes, the radiated energy is found not to be
a monotonically increasing function of initial separation; there is a
local radiated-energy maximum at $r_0\approx4.5M$. 
The present results, along with results for infall from infinity,
provide a complete catalog of waveforms and spectra for particle
infall. We give a representative sample from that catalog and an
interesting observation: Unlike the simple spectra for other head-on
collisions (either of particle and hole, or of equal mass holes) the
spectra for $\infty>r_0>\sim5M$ show a series of evenly spaced bumps. A
simple explanation is given for this.
Lastly, our energy vs. $r_0$ results are compared with approximation
methods used elsewhere, for small and for large initial separation.
\end{abstract}

\section{Introduction and Overview}
Among the earliest calculations of astrophysical sources of
gravitational radiation is the ``DRPP'' calculation
\cite{drpp}\cite{petrich} of radiation emitted when a particle,
starting from
rest at infinity, falls into a nonspinning black hole. In this work,
the mass of the particle is treated as a perturbation of the
Schwarzschild spacetime and the Einstein equations, to first order in
this perturbation, are organized into the Zerilli
equation\cite{zerilli,moncrief}, a single linear wave equation. This
work was later elaborated  by Ruffini and Ferrari\cite{ruff-ferr}
who considered the head-on plunge of a particle which starts
with nonzero inward velocity at spatial infinity.

There is now a renewed interest in such calculations. The advent of
laser interferometric gravity wave detectors\cite{ligo} has directed
attention at black hole collisions as one of the most plausible, and
surely the most fascinating, sources of detectable waves. But the
generation of strong waves is a process without simplifying symmetries
and one which involves strong field interactions. The challenge of
finding energies and waveforms generated in these processes helped to
spur the recent effort in numerical relativity and, in particular, in
the Binary Black Hole supercomputing Grand
Challenge\cite{grandchallenge}.
Because of the difficulty of direct supercomputer solutions,
approximate methods, as checks and guides, are very valuable. The
``particle limit'' in which the mass of one of the holes is very
small, is in this category, and models of holes in binary orbits are a
fundamental tool. 

We are interested here in using a particle model to help in the
understanding of numerical relativity results for collisions of holes.
To date those computations have been limited to head on, axisymmetric
collisions\cite{annprl,ast}. The codes have also been limited in the
time for which they can evolve solutions. Infall from very large
distances cannot at present be evolved; computations start from finite
separation of the holes, so direct comparison with the DRPP results
cannot be made.  We present here an extension of the DRPP calculation
to the case of infall from finite radius.

The extension is not at all straightforward. If the particle starts at
infinity, the initial value data can be taken as zero, so that the
spacetime is unperturbed Schwarzschild until the gravitational
influence of the particle is felt. If the particle starts from finite
radius, the specification of compatible initial value data cannot be
avoided. Understanding the influence of the initial data is, in fact,
a major motivation for the present work. Supercomputer evolution will
be limited, for the foreseeable future, to evolving the interaction of
holes for short times. The starting point for these evolutions will be
initial data from some approximation method (e.g., a post-Newtonian
expansion). The way to make the bridge from the approximation to the
strong field realm of supercomputing requires that we understand what
the crucial features are of the initial value data. It seems to us
that the particle limit provides a very useful probe of this issue.

The remainder of the paper is organized as follows. In Sec.~II we
start by presenting the basic mathematical approach. Like the DRPP
calculation, our approach uses frequency components. A Laplace
transform converts the linearized Einstein equation for each
$\ell$-pole, from a partial differential equation in Schwarzschild
coordinates $r,t$ to a set of ordinary differential equations in $r$.
The Laplace transform treats the initial data mathematically as if it
were a source, and helps with insight into the relation of the initial
value data and the stress-energy source.  Two points in particular
should be noted in Sec.~II. The first concerns the linearized wave
equation. The equation presented by Zerilli was formulated in the
Regge-Wheeler gauge\cite{rw}, a specific first order coordinate
choice. The Zerilli equation was presented through a Fourier transform
and has no direct equivalent as a partial differential equation. This
is not a fundamental obstacle, but it does complicate the relationship
of the Zerilli wave function and the initial data. To avoid these
problems we use the Moncrief formulation of the
problem\cite{moncrief}. Moncrief's wave function is constructed
explicitly from initial data and is equivalent to the Zerilli wave
function except for source-like terms.  The second point to notice has
to do with conventions. A variety of conventions has been used for
normalizing the Zerilli (or Moncrief) functions and their
transforms. In Sec.~II, we present a definition of all our conventions
and their relationship to those used elsewhere.

Aspects of the numerical approach, a Green function integral, are
discussed in Sec.~III. A major issue is the need for highly accurate
evaluation of the Green function integrals in order to preclude
numerical errors from giving rise to an erroneous initial burst of
radiation.  Numerical results are presented in Sec.~IV for energy,
spectra, and waveforms. Plots are given for infall starting from rest
at radii ranging from $r_0/2M=1.1$ to $r_0/2M=15$. These are combined
with results for infall from infinity to give a complete picture of
the dependence of the radiation on the initial physical conditions.
The numerical results are discussed in Sec.~V and are compared with
predictions based on other results and approximations, especially
those of the close-limit approximation.

Throughout the rest of the paper we use $c=G=1$ units, the signature
$-+++$, and other conventions of the Misner, Thorne and Wheeler
textbook\cite{mtw}.

\section{Mathematical approach} 
\subsection{Variables and conventions}

For the straight-line plunge of a
particle into a nonspinning hole the perturbations of the
Schwarzschild spacetime are all even-parity.  We take the particle to
be moving along the $z$ axis, so that the perturbations are not
dependent on the azimuthal angle $\phi$, To describe the perturbations
we use the notation of Regge and Wheeler\cite{rw}, here specialized to
the axisymmetric $\ell$-pole case ($m=0$):
\begin{eqnarray}\label{rwform}
ds^2=ds^2_0+(1-2M/r)(H_0Y_{\ell 0})dt^2+ (1-2M/r)^{-1}(H_2Y_{\ell
0})dr^2\nonumber\\ +r^2(KY_{\ell 0}+G\partial^2Y_{\ell
0}/\partial\theta^2)d\theta^2 +r^2(\sin^2\theta KY_{\ell
0}+G\sin\theta\cos\theta\partial Y_{\ell 0}/\partial\theta
)d\phi^2\nonumber\\ +2H_1Y_{\ell 0}dtdr +2h_0(\partial Y_{\ell
0}/\partial\theta) dtd\theta +2h_1(\partial Y_{\ell 0}/\partial\theta)
drd\theta
\end{eqnarray}
where $ds^2_0$ is the unperturbed line element for a Schwarzschild
spacetime of mass $M$, where $H_0,H_1,H_2,h_0,h_1,K$ and $G$ are
functions of $r,t$, and where $Y_{\ell 0}(\theta)$ are the $m=0$
spherical harmonics.  We then follow the prescription given by
Moncrief\cite{moncrief} in his eqs.  (5.8) -- (5.12) and (5.27) to
arrive at a wave function.  Rather than choose the normalization of
Moncrief's ``$Q$'' we choose a normalization closely related to
Zerilli's\cite{zerilli} wave function.  In the Regge-Wheeler\cite{rw}
gauge ($G=h_0=h_1=0$) this ``Moncrief-Zerilli'' function is
\begin{equation}\label{psidef}
  \psi(r,t)=\frac{r}{{\lambda}+1}\left[
    K+\frac{r-2M}{{\lambda}r+3M}\left\{ H_2-r\partial K/\partial r
    \right\} \right]
\end{equation}
where we have used Zerilli's notation
\begin{equation}{\lambda}\equiv(\ell+2)(\ell-1)/2\ .
\end{equation}
We now give the relationship of this variable, which we shall use
throughout the paper, to the following wavefunctions: (i) $Q$
appearing in Moncrief's\cite{moncrief} eq.~(5.27) (ii) $\psi_{\rm
pert}$ and $\psi_{\rm num}$ appearing in Ref.~\cite{annetal}.  For
$\ell=2$ the wave function appearing as $\tilde{\psi}$ in eq.~(II-31)
of Cunningham, Price and Moncrief\cite{cmp}, and $\psi$ defined in
eq.~(13) of Price and Pullin\cite{pp}, agree with $\psi_{\rm
pert}$. The relations are:
\begin{eqnarray}
&\psi=2Q/[\ell(\ell+1)]\nonumber\\
 &=2[(\ell-2)!/(\ell+2)!]\,\psi_{\rm pert}\\
&=\sqrt{2[(\ell-2)!/(\ell+2)!]}\,\psi_{\rm num}\ .\nonumber
\end{eqnarray}

Zerilli\cite{zerilli} derives his equation only in the frequency
domain, so no direct comparison can be made with functions of
$r,t$. It is, however, straightforward to use the same steps as
Zerilli in the Regge-Wheeler gauge. but confined to the $r,t$ domain.
It turns out that to decouple the even-parity variables in the
perturbed Einstein equations, one needs to take an extra time
derivative. One can then define the variable $\psi_{\rm Zer}(r,t)$ by
\begin{equation}\label{zerillidef}
\dot{\psi}_{\rm Zer}=\frac{r^2\dot{K}-(r-2M)H_1
}{{\lambda}r+3M}\ ,
\end{equation}
where the dot indicates a derivative with respect to $t$.  The Fourier
transform of this variable, divided by $-i\omega$ is the function
defined in Zerilli's papers. For vacuum perturbations $\dot{\psi}_{\rm
Zer}=\dot{\psi}$, so the Zerilli wave function, in vacuum, can be
chosen to agree with our $\psi$.

By computing the stress-energy pseudotensor in a radiation gauge, we
find that radiated power per solid angle is
\begin{equation}
\frac{d{\rm Power}}{d\Omega}=\frac{1}{64\pi}
\left[\frac{\partial\psi}{\partial u}\right]^2
\left(
\cot\theta\frac{\partial}{\partial\theta}Y_{\ell0}
-\frac{\partial^2}{\partial\theta^2}Y_{\ell0}
\right)^2
\end{equation}
(where $u$ is retarded time), so that the gravitational wave power,
integrated over all angles, is
\begin{equation}\label{power}
{\rm Power}=\frac{1}{64\pi}\frac{(\ell+2)!}{(\ell-2)!}
\left[\frac{\partial\psi}{\partial u}\right]^2\ .
\end{equation}

Our wave function $\psi$ satisfies a second-order
wave equation with a source term,
\begin{equation}\label{rtzerilli}
-\frac{\partial^2\psi}{\partial t^2}
+\frac{\partial^2\psi}{\partial r*^2}-V_{\ell}(r)\psi=
{\cal S}_\ell (r,t)\ .
\end{equation}
Here $r^*\equiv r+2M\ln(r/2M-1)$ is the Regge-Wheeler\cite{rw}
``tortoise'' coordinate, $V_\ell$ is the Zerilli potential
\begin{equation}\label{zpotential}
V_\ell(r)=\left(1-\frac{2M}{r}\right)
\frac{2\lambda^2(\lambda+1)r^3+6\lambda^2Mr^2+18\lambda M^2r+18M^3
}{r^3(\lambda r+3M)^2 }\ ,
\end{equation}
and ${\cal S}_\ell (r,t)$ is the source term.

We are interested in the case in which the source is a point particle,
of mass $m_0$, following a spacetime trajectory given by $t=T(\tau)$,
and $r=R(\tau)$, where $\tau$ is proper time along the particle world
line,
\begin{equation}
T^{\mu\nu}=(m_0/U^0)U^\mu U^\nu \delta(r-R)\delta^2(\Omega)/r^2\ .
\end{equation}
Here $U^\mu\equiv dx^\mu/d\tau$ is the particle 4-velocity, and the
0-component is $U^0=\varepsilon_0/(1-2M/r)$, where
$\varepsilon_0=-U_0$, the ``energy-at-infinity per unit particle
mass,'' is a constant of motion that takes the value $\sqrt{1-2M/r_0}$
for infall from rest at $r_0$.  The two dimensional delta function
$\delta^2(\Omega)$ gives the angular location of the particle
trajectory
\begin{equation}
\delta^2(\Omega)=\sum_{\ell,m}Y_{\ell m}(\theta,\phi)Y_{\ell m}^*
(\theta,\phi)=\sum_\ell Y_{\ell0}(\theta)\sqrt{(2\ell+1)/4\pi}\ ,
\end{equation}
with the last expression applying for infall along the positive $z$
axis.

We now need to write the  the
Einstein ($G_{\mu\nu}$) and Ricci ($R_{\mu\nu}$) tensors, in a tensor
spherical harmonic decomposition like that in (\ref{rwform}). The
components we shall need are:
\begin{equation}
G_{00}\equiv \sum_{\ell}G44 Y_{\ell0}\ \ \ R_{00}\equiv
\sum_{\ell}R44 Y_{\ell0}
\ \ 
R_{\theta\theta}\equiv \sum_{\ell}( R22K Y_{\ell0}
+R22G\partial^2Y_{\ell0}/\partial\theta^2)\ .
\end{equation}
(Here, and below, we omit $\ell$ indices where the meaning of the
symbols -- such as $G44$ -- is clear from context.)  By rearranging the
expressions for the perturbed Einstein and Ricci components, we find
that the source term in (\ref{rtzerilli}), for an axisymmetric
problem, is
\begin{eqnarray}\label{srt}
  {\cal S}_\ell (r,t)=\frac{2(1-2M/r)}{r(\lambda+1)(\lambda r+3M)}
  \left[r^2\left(1-\frac{2M}{r}\right)\frac{\partial}{\partial
      r}R22K\right.\nonumber\\
\left.
-(\lambda r+M)R22K-\frac{r^4\lambda}{\lambda r+3M}G44
+r^3R44\right]\ .
\end{eqnarray}
The Einstein equations can now be used to replace the perturbed
Ricci and Einstein components with components of the particle
stress energy 
\begin{eqnarray}
&G44=\kappa U^0(1-2M/r)^2\delta(r-R)/r^2\nonumber\\
&R44=\kappa\
\left[(1-2M/r)^2
 U^0
-\frac{1}{2 U^0}(1-2M/r)
\right]\delta(r-R)/r^2\\
&R22K=\kappa\left[\delta(r-R)/2 U^0
\right]\ ,\nonumber
\end{eqnarray}
where
\begin{equation}
\kappa\equiv8\pi m_0\sqrt{(2\ell+1)/4\pi}\ .
\end{equation}
When these are used in (\ref{srt}) the result is
\begin{eqnarray}\label{rtsource}
{\cal S}_\ell (r,t)=-\frac{2(1-2M/r)\kappa
}{r(\lambda+1)(\lambda r+3M)
}\left[-r^2(1-2M/r)\frac{1}{2 U^0}\delta'(r-R)\right.\nonumber\\
+\left.\left\{\frac{r(\lambda+1)-M}{2 U^0}-\frac{3M U^0 r(1-2M/r)^2
}{r\lambda+3M}\right\}\delta(r-R)
\right]\ .
\end{eqnarray}
When the source term is included, the relation between our wave function
$\psi$, and the Zerilli wave function of (\ref{zerillidef})
becomes
\begin{equation}
  \dot{\psi}=\dot{\psi}_{\rm Zer}-\frac{\kappa U^0(r-2M)}{
    (\lambda+1)(\lambda r+3M) }\,\frac{dR}{dt}\,\delta(r-R)\ .
\end{equation}

\subsection{Initial data}
For comparison with established results of numerical relativity we
want initial data representing an initially stationary spacetime, so
we take $\dot{\psi}_0$, the initial time derivative of $\psi$, to be
zero.  To determine our $\psi_0$, the initial value of $\psi$, we
choose an initial 3-geometry with the conformally flat form that is
used in numerical relativity\cite{bl,ap2},
\begin{equation}\label{conflfrm}
ds^2=\Phi(\bar r,\theta)^4\left[d\bar r^2+\bar r^2
(d\theta^2+\sin^2\theta d\phi^2)\right]\ ,
\end{equation}
where $\Phi$ satisfies a flat-space Laplace equation $\nabla^2\Phi=0$.
We choose $\Phi$ to represent a throat of mass $m_1$ on the $z$ axis
at $z=z_1$ and a throat of mass $m_2$ at $z=z_2$. The simplest
such solution has the form
\begin{equation}\label{blsoln}
\Phi=1+\frac{1}{2}
\left(\frac{m_1}{\sqrt{\bar r^2\sin^2{\theta}+(\bar r\cos{\theta}-z_1)^2}}
+\frac{m_2}{\sqrt{\bar r^2\sin^2{\theta}+(\bar r
\cos{\theta}-z_2)^2}}\right)\ .
\end{equation}
We now identify $M\equiv m_1+m_2$, 
and choose $z_1=-z_2(m_2/m_1)$ so that the dipole moment vanishes for
$\bar r>z_2$.
We treat the mass ratio
$m_2/M$ as small and keep terms in (\ref{blsoln}) only to first order
in this ratio.  We ignore the nonradiative $\ell=0,1$ perturbations
and write $\Phi$ as
\begin{equation}
\Phi=1+\frac{M}{2\bar r}+\frac{m_2}{2\bar r}\sum_{\ell=2,3,\dots}
{\cal F}_\ell(\bar r)
P_\ell(\cos{\theta})
\end{equation}
where
\begin{equation}\label{Fform}
{\cal F}_\ell(\bar r)=\left\{\begin{array}{ll}
(z_2/\bar r)^\ell &\mbox{if $\bar r>z_2$}\\
(\bar r/z_2)^{\ell+1} &\mbox{if $\bar r<z_2$}
\end{array}
\right.\ .
\end{equation}

We next change radial variables, from isotropic-like $\bar r$ to a
Schwarzschild-like coordinate $r$, with the transformation
\begin{equation}\label{eq:Rdef}
\bar r=\left(\sqrt{r} +\sqrt{r-2M}\right)^2/4\ .
\end{equation}
When this is used in (\ref{conflfrm}), and terms higher order in $m_2$
are omitted, the result is
\begin{equation}\label{ivmetric}
ds^2=\left(1+\frac{2m_2/\bar r}{1+M/2\bar r}\sum_{\ell=2,3,\dots}
{\cal F}_\ell(\bar r)
P_\ell(\cos{\theta})
\right)\left[\frac{dr^2}{1-2M/r}+
r^2(d\theta^2+\sin^2\theta
d\phi^2)
\right]\ .
\end{equation}
{}From this we can infer that the initial value perturbtations in the
Regge-Wheeler notation of (\ref{rwform}) are $G=h_1=0$ and
\begin{equation}\label{initK}
K=H_2=\frac{2m_2/\bar r}{1+M/2\bar r}
{\cal F}_\ell(\bar r)
\sqrt{\frac{4\pi}{2\ell+1}}\ .
\end{equation}

We must now complete the identification of this conformally flat
solution with the point particle solution. Clearly the location
$\bar r=z_2$ of the perturbative throat must be set to
\begin{equation}\label{z2eq}
z_2\equiv \bar r_0=\left(\sqrt{r_0} +\sqrt{r_0-2M}\right)^2/4\ .
\end{equation}
The mass $m_2$ of the throat must {\em not} be set to the mass $m_0$
of the particle. The particle mass $m_0$ can be viewed as the ``bare''
mass of the particle, the mass that is measured very close to the
location of the particle. On the other hand, $m_2$ is a formal
parameter of the initial value solution; since the total ADM mass for
(\ref{blsoln}) is $m_1+m_2$, gravitational binding energy is included
in $m_2$. We give here two arguments for the correct relationship.
The first is based on the definition, given by Brill and
Lindquist\cite{bl}, of ``bare'' mass for a solution with the form
(\ref{blsoln}), a mass exclusive of binding energy.  To find an
expression for this bare mass they look at the extension of the
geometry through the perturbative throat into an asymptotically flat
universe; the bare mass of the perturbative throat is the mass
measured at infinity in that universe. From eq.~(13) of
Ref.~\cite{bl} the bare mass of
the perturbative throat (to first order in $m_2/M$) is 
\begin{equation}
m_2^{\rm bare}=m_2(1+M/2\bar r_0)\ .
\end{equation}
We identify the bare mass with the particle mass $m_0$
and conclude
\begin{equation}\label{massreln}
m_2=m_0\left(1+\frac{M}{2\bar r_0}\right)^{-1}
={\textstyle\frac{1}{2}}m_0\left(1+\sqrt{1-\frac{2M}{r_0}}\right)\ .
\end{equation}
An independent way of finding the relationship is to look at the
perturbed hamiltonian constraint. For a conformally flat ($H_0=H_2=K$)
initial 3-geometry [equivalent to eq. (C7a) of Ref.~\cite{zerilli2}]
this gives us
\begin{eqnarray}\label{constraint}
\left(
1-\frac{2M}{r}
\right)^2 \frac{\partial^2K}{\partial r^2}
+\left(
1-\frac{2M}{r}
\right) 
\left(
2-\frac{3M}{r}
\right) \frac{1}{r}
\frac{\partial K}{\partial r}
-\left(
1-\frac{2M}{r}
\right) \frac{\ell(\ell+1)}{r^2}K
=-G44\nonumber\\ = -8\pi m_0\sqrt{\frac{2\ell+1}{4\pi}} U^0\left(
1-\frac{2M}{r}
\right)^2\frac{1}{r^2}\delta(r-r_0)\ .
\end{eqnarray}
By integrating across the discontinuity, and by using
(\ref{Fform}),(\ref{eq:Rdef}), and (\ref{initK}),
we find
\begin{eqnarray}\label{jumpcond}
\Delta K,_r\equiv dK/dr|_{r=r_0^+}-dK/dr|_{r=r_0^-}\nonumber\\
=-8\pi m_2\frac{\sqrt{(2\ell+1)/4\pi}}{\bar r_0^{1/2}
r_0^{3/2}\sqrt{1-2M/r_0}}\\
 =-8\pi m_0\nonumber
\frac{\sqrt{(2\ell+1)/4\pi}}{r_0^2\sqrt{{1-2M/r_0}}}\ ,
\end{eqnarray}
which gives us the same relationship as in (\ref{massreln}).  This
second derivation of the mass relationship is instructive.  If we
choose a conformally flat initial geometry then all the information
about the perturbations of that geometry is contained in the single
function $K(r)$. But (\ref{constraint}) completely fixes $K(r)$, so
the initial geometry has no freedom. (Alternatively: To choose
different initial data we would have to introduce nonconformally-flat
terms.)
This is to be contrasted with the case of nonperturbative throats,
where a variety of choices is possible for conformally flat initial
data.

\subsection{Laplace transforms}
We now define the Laplace transform $\Psi$
of $\psi$ to be 
\begin{equation}\label{xform}
\Psi(r,\omega)\equiv\int_0^\infty e^{i\omega t}\psi(r,t)\,dt\ .
\end{equation}
We take $\psi$ to vanish for $t<0$, which means that $\Psi(r,\omega)$
must be analytic in the upper half of the complex $\omega$ plane.
At large $r$, for outgoing waves, $\psi(r,t)$ is a 
function only of $t-r^*$ so $\Psi$ takes the form
\begin{equation}\label{ampintro}
\Psi(r,\omega)\rightarrow A(\omega)e^{i\omega r^*}\ .
\end{equation}
We call $A$ the amplitude of the outgoing radiation.
Since the outgoing radiation $\psi(r,t)$ is pure 
real, the amplitude satisfies the crossing relation
$A(-\omega)=A^*(\omega)$.

The waveform for outgoing radiation, as a function of retarded time
$u\equiv t-r^*$ can be found by the inverse transform 
\begin{equation}\label{inverse}
\psi(u)=\frac{1}{2\pi}\int_{-\infty}^{+\infty}A(w)e^{-i\omega u}d\omega
={\rm Re}\left[\frac{1}{\pi}\int_0^\infty A(w)e^{-i\omega u}d\omega
\right]\ .
\end{equation}
For a particle falling in from infinity the waveform extends to
$u\rightarrow-\infty$, and there is no initial data to deal with.  In
this case $A(\omega)$ is to be interpreted as a Fourier, rather than
Laplace, transform. If the particle has $\varepsilon_0>1$, then the
waveform does not vanish at $u\rightarrow-\infty$, and the transform
exists only for $\omega$ in the lower half plane, and will have a pole
at $\omega=0$. For the inverse transform (the first integral in
(\ref{inverse}) the contour should be interpreted as going below the
real $\omega$ axis, so the integral is the equivalent of the Cauchy
principal value plus half the contribution of the residue of the pole
at $\omega=0$,
\begin{equation}
  \psi(u) ={\rm
    Re}\left[\lim_{\epsilon\rightarrow0}\frac{1}{\pi}\int_\epsilon^\infty
    A(w)e^{-i\omega
    u}d\omega\right]+\frac{i}{2}\lim_{\omega\rightarrow0}\left\{
\omega A(\omega)
\right\}\ .
\end{equation}

{}From Parseval's theorem and (\ref{power}), we
have that the radiated energy at Scri+ is
\begin{eqnarray}\label{wenergy}
  {\rm Energy}=\frac{1}{64\pi}\frac{ (\ell+2)!
    }{(\ell-2)!}\int_{-\infty}^\infty
  \left(\frac{d\psi}{du}\right)^2\,dt\nonumber\\ 
  =\frac{1}{128\pi^2}\frac{ (\ell+2)!
    }{(\ell-2)!}\int_{-\infty}^\infty
  \omega^2|A(\omega)|^2\,d\omega\nonumber\\ =\frac{1}{64\pi^2}\frac{
    (\ell+2)!  }{(\ell-2)!}\int_0^\infty
  \omega^2|A(\omega)|^2\,d\omega\ .
\end{eqnarray}
The energy spectrum is therefore given by
\begin{equation}\label{spectrum}
\frac{d\rm Energy}{d\omega}=\frac{1}{64\pi^2}\frac{
    (\ell+2)!  }{(\ell-2)!}  \omega^2|A(\omega)|^2\ .
\end{equation}

We now multiply (\ref{rtzerilli})
by $e^{i\omega t}$, integrate from $t=0$ to $\infty$, and integrate
by parts in the integral involving $\partial^2\psi/\partial t^2$. The 
result is an ordinary differential equation for $\Psi$:
\begin{equation}\label{basiceq}
  \frac{\partial^2\Psi}{ \partial r*^2}
  +\left[\omega^2-V_\ell(r)\right]\Psi=-\dot{\psi}_0(r)
+i\omega\psi_0(r)+S(r,\omega)\ ,
\end{equation}
where $\psi_0(r)$ is the initial value of $\psi(t,r)$, and
$\dot{\psi}_0(r)$ is the initial value of $\dot{\psi}(t,r)$, and
the source term $S$ is defined as 
\begin{equation}\label{srctrans}
S(r,\omega)\equiv\int_0^\infty e^{i\omega t}{\cal S}_\ell (r,t)dt\ .
\end{equation}
For the particle source   
we substitute (\ref{rtsource}) into  (\ref{srctrans}), to get
\begin{eqnarray}\label{wrsource}
S(r,\omega)=\frac{2(1-2M/r)\kappa}{r(\lambda+1)(\lambda r+3M)}
\left[-r^2(1-2M/r)
\left\{
\frac{1}{|\dot{R}|}\frac{d}{dt}
\left(\frac{e^{i\omega t}}{2 U^0\dot{R}}\right)\right\}\right.\nonumber\\
+\frac{1}{|\dot{R}|}\left.\left\{\frac{r(\lambda+1)-M}{2 U^0}-
\frac{3M U^0 r(1-2M/r)^2}{r\lambda+3M}\right\}e^{i\omega t}
\right]_{T(r)}\ .
\end{eqnarray}
The subscript ``$T(r)$'' indicates that functions of time, such as
$\dot{R}\equiv dR/dt,  U^0, e^{i\omega t}$ are to be evaluated at the
value of $t=T(r)$ where $T=T(\tau)$ with $r=R(\tau)$. For infall from
rest at radius $r_0$, this gives $t$ as a function of $r$ through
\begin{eqnarray}\label{time}
T(r)&=& \varepsilon_0\left({r_0\over2M}\right)
\left({r\over2M}\right)^{1/2}
\sqrt{1-{r\over r_0}}
\nonumber\\
&&+(1+{4M\over r_0})\left({r_0\over2M}\right)^{3/2}\varepsilon_0
\arctan\left[\sqrt{{r_0\over r}-1}\right]\nonumber\\
&&+2\,{\rm arctanh}\left[\varepsilon_0^{-1}
\sqrt{{2M\over r}-{2M\over r_0}}\right]\ .
\end{eqnarray}
The result in (\ref{wrsource}) for $S(r,\omega)$,
however, is not valid for $r=r_0$. When $r=r_0$ the delta functions in
(\ref{rtsource}), as functions of $t$, have vanishing arguments at
$t=0$, the endpoint of $t$ integration in (\ref{srctrans}), so the
integration is not well defined. The way to deal with this will be
explained in the next section.

\section{Computational implementation} 
\subsection{Green function formal solution}

We start by rewriting (\ref{basiceq}) in the form
\begin{equation}\label{totaleq}
  \frac{\partial^2\Psi}{ \partial r^{*2}}
  +\left[\omega^2-V_\ell(r)\right]\Psi=S_{\rm tot}(r,\omega)\ ,
\end{equation} 
where $S_{\rm tot}$ is the complete right hand
side of (\ref{basiceq}), including both the stress energy term and the
initial value terms,
\begin{equation}\label{termsagain}
S_{\rm tot}(\omega,r)=-\dot{\psi}_0(r)+i\omega\psi_0(r)+S(r,\omega)\ .
\end{equation}
This equation is to be solved for the boundary conditions of ingoing
waves at the horizon, and outgoing waves at spatial infinity:
$\Psi\rightarrow e^{-i\omega r^*}$ for $r^*\rightarrow-\infty$, and
$\Psi\rightarrow e^{i\omega r^*}$ for $r^*\rightarrow+\infty$.  The
Green function solution is found in the usual way. We define
$y_L(r^*,\omega)$ and $y_R(r^*,\omega)$ as the homogeneous solutions of
(\ref{totaleq}) with asymptotic forms
\begin{eqnarray}
  y_L(r^*,\omega)\stackrel{r^*\rightarrow-\infty}{\longrightarrow}
  e^{-i\omega r^*}\nonumber\\
  y_R(r^*,\omega)\stackrel{r^*\rightarrow\infty}{\longrightarrow}
  e^{i\omega r^*}\ .
\end{eqnarray}
We define the Wronskian of the homogeneous solutions,
an $r^*$-independent constant, to be 
\begin{equation}\label{wronsk}
  W(\omega)\equiv
  y_L\frac{d}{dr^*}y_R-y_R\frac{d}{dr^*}y_L\ .
\end{equation}
With the above definitions, the Green function solution is written
\begin{equation}\label{gfsoln}
  \Psi(r,\omega)=\frac{1}{W(\omega)}\left[
    y_R(r^*,\omega)\int_{-\infty}^{r^*}S_{\rm tot}(\tilde{r},\omega)
    y_L(\tilde{r^*},\omega)d\tilde{r^*} +
    y_L(r^*,\omega)\int_{r^*}^{\infty}S_{\rm tot}(\tilde{r},\omega)
    y_R(\tilde{r^*},\omega)d\tilde{r^*} \right]\ .
\end{equation}
In the limit of large $r^*$ this gives us
\begin{equation}\label{ampgreen}
A(\omega)=\frac{1}{W(\omega)}
\int_{-\infty}^{\infty}S_{\rm tot}(r,\omega)
    y_L(r^*,\omega)dr^*\ .
\end{equation}

\subsection{Evaluation of the Green function integral}
The first two terms in (\ref{termsagain}) are straightforward to
integrate in (\ref{ampgreen}), but the stress energy source term
$S(r,\omega)$ cannot be evaluated at $r_0$. [The expression in
(\ref{wrsource}) formally diverges as $(r-r_0)^{-3/2}$ and hence
cannot be used in (\ref{ampgreen}).]
To make sense of this we start by writing the source in
(\ref{rtsource}) as
\begin{equation}
S(r,t)=F(r,t)\delta'(r-R[t]))+G(r,t)\delta(r-R[t]))\ ,
\end{equation}
where $F$ and $G$ contain no delta functions. The troublesome 
part of the Green function integral can then be written as
\begin{equation}
  \int_{-\infty}^{+\infty}S(r,\omega)y_L(r^*,\omega)\ dr^*={\cal
    I}_1+{\cal I}_2\ ,
\end{equation}
where ${\cal I}_1$ is the integral involving $G$ and ${\cal I}_2$
involves $F$. The first of these is
\begin{eqnarray}
  {\cal I}_1&=\int_{0}^{+\infty}e^{i\omega t}dt\int_{-\infty}^\infty
   y_L(r^*,\omega)dr^* G(r,t)\delta(r-R)\nonumber\\
   &=\int_0^\infty e^{i\omega t}dt
   \int_{2M}^{\infty}y_L(r^*,\omega)
   G(r,t)\delta(r-R)\frac{dr}{1-2M/r}\nonumber\\ &=\int_0^\infty
   e^{i\omega t}y_L(r^*(t),\omega)
   G(R(t),t)\,dt/(1-2M/R(t))\nonumber\\
   &=-\int_{-\infty}^{r_0^*}e^{i\omega
   T(r)}y_L(r^*,\omega)G(r,T(r))dr^*/\dot{R}\ .
\end{eqnarray}
In the final integral, the factor $1/\dot{R}\sim(r-r_0)^{-1/2}$
diverges but is integrable.
A similar set of transformations is now applied to ${\cal I}_2$,
\begin{eqnarray}
  {\cal I}_2&=&\int_{0}^{+\infty}e^{i\omega t}dt
\int_{-\infty}^\infty y_L(r^*,\omega)
   F(r,t)\delta'(r-R)\,dr^*\nonumber\\ 
&=&\int_0^\infty e^{i\omega t}dt
  \int_{2M}^{\infty}y_L(r^*,\omega) 
  F(r,t)\delta'(r-R)\frac{dr}{1-2M/r}\nonumber\\ 
&=&
-\int_0^\infty e^{i\omega t}dt 
\left[y_L(r^*(t),\omega)\frac{\partial}{\partial r}\left(\frac{F}{1-2M/r}
\right)
+\frac{F}{(1-2M/r)^2}\frac{\partial}{\partial r^*}\left(y_L(r^*(t),\omega)
\right)
\right]_{r=R(t)}
\\
&=&\int_{-\infty}^{r_0^*}e^{i\omega
T(r)}\left[\left(1-\frac{2M}{r}\right)
\frac{\partial}{\partial r}\left(\frac{F}{1-2M/r}
\right)y_L(r^*,\omega)+\left(\frac{F}{1-2M/r}
\right)
\frac{\partial}{\partial r}\left(y_L(r^*,\omega)\right)
\right]\frac{dr^*}{\dot{R}}\ .\nonumber
\end{eqnarray}

To evaluate these explicitly we need the fact that for free fall from
rest at $r_0$ 
\begin{equation}
\dot{R}=-\left(1-\frac{2M}{r}
\right)\sqrt{\frac{2M/r-2M/r_0}{1-2M/r_0}}\ .
\end{equation}  
We
can now use the explicit expressions for $F(r,t),G(r,t)$ from
(\ref{rtsource}) to write
\begin{eqnarray}\label{I1andI2}
{\cal I}_1+{\cal I}_2=-\frac{\kappa}{\lambda+1}
\int_{-\infty}^{r_0^*}\frac{e^{i\omega T(r)}dr^*}{
\sqrt{2M/r-2M/r_0}}
\frac{(1-2M/r)}
{(\lambda r+3M)}
\left[r\frac{\partial}{\partial r^*}y_L\right.\nonumber\\
\left.+y_L\left\{\lambda+1-\frac{M}{r}+\frac{(2\lambda-3+12M/r_0)M}
{\lambda r+3M}\right\}
\right]\ .
\end{eqnarray}
{}From (\ref{ampgreen}), (\ref{termsagain}), and the definitions of
${\cal I}_1+{\cal I}_2$, we have that
\begin{equation}\label{sofar}
A(\omega)=\frac{1}{W(\omega)}\left[{\cal I}_1+{\cal I}_2
+i\omega\int_{-\infty}^{\infty}\psi_0(r)y_L(r^*,\omega)\,dr^*
-\int_{-\infty}^{\infty}\dot{\psi}_0(r)y_L(r^*,\omega)\,dr^*\right]\ .
\end{equation}
Since we are considering an initially stationary problem we have
$\dot{\psi}_0=0$. The initial data for $\psi$ comes from putting
(\ref{initK}) in 
(\ref{psidef}):
\begin{eqnarray}\label{psi0}
\psi_0=\frac{2m_2}{\lambda+1}
\frac{
\sqrt{4\pi/(2\ell+1)}}
{1+M/2\bar r}
\frac{r}{\lambda r+3M}
\times\hspace*{.6in}\nonumber\\
\left[
\left( (\lambda+1)r+M-r\sqrt{1-2M/r}\frac{M/2\bar r}{1+M/2\bar r}
\right)\right.
\left\{\begin{array}{c}
\bar r_0^\ell/\bar r^{\ell+1}\\\bar r^\ell/\bar r_0^{\ell+1}
\end{array}
\right\}
\\
\left.
-r\sqrt{ 1-\frac{2M}{r}}\bar r
\left\{\begin{array}{c}
-(\ell+1)\bar r_0^\ell/\bar r^{\ell+2}\\\ell \bar r^{\ell-1}/
\bar r_0^{\ell+1}
\end{array}
\right\}\right]\ ,\hspace*{.4in}\nonumber
\end{eqnarray}
where the upper expressions apply in the case $\bar r>\bar r_0$,
and the lower for $\bar r<\bar r_0$.

It should be noted that $\psi_0$ approaches a nonzero constant, and
$y_L(r^*,\omega)\rightarrow e^{-i\omega r^*}$, as $r\rightarrow2M$, so
the integral over $\psi_0$, in (\ref{sofar}) is improper at large
negative $r^*$. We must recall that we are really computing $A$
in the upper half of the complex $\omega$ plane. To deal with this
computationally we recast the integral in (\ref{sofar}) into the form
\begin{equation}\label{trick}
\int_{-\infty}^{\infty}\psi_0(r)y_L(r^*,\omega)\,dr^*
\rightarrow 
i\omega^{-1}\psi_0(2M)e^{-i\omega r^*_{\rm start}}+
\int_{r_{\rm start}^*}^{\infty}\psi_0(r)y_L(r^*,\omega)\,dr^*\ .
\end{equation}
where $\psi_0(2M)$ is the limit of $\psi_0$ at $r=2M$. The value of
$r^*_{\rm start}$ must be chosen such that there is negligible
variation of $\psi$ between $r^*_{\rm start}$ and the horizon.

\subsection{Numerical method}
The first step in the solution is to determine the Wronskian in
(\ref{wronsk}).  We denote the form of $y_L$ at large $r$ by
\begin{equation}\label{asympt}
y_L\stackrel{r\rightarrow\infty}{\longrightarrow}
\alpha(\omega)e^{i\omega r^*}+
\beta(\omega)e^{-i\omega r^*}\ .
\end{equation}
We find $\beta(\omega)$ by solving (\ref{totaleq}) with the righthand
side set to zero, and with the starting condition $y_L=e^{-i\omega
r^*}$ imposed at a large negative value of $r^*$.  A fourth-order
Runge-Kutta routine is used to integrate $y_L$ out to large values of
$r^*$ where it is matched to approximate forms of the asymptotic
solution. In practice, good accuracy was difficult to achieve with the
asymptotic form in (\ref{asympt}) and asymptotic solutions one order
higher in $1/\omega r$ were used.  From $\beta$, the Wronskian 
follows immediately:
\begin{equation}
W(\omega)=2i\omega\beta(\omega)\ .
\end{equation}
With $\beta$ in hand, with $\dot{\psi}_0=0$, and with the substitution
in (\ref{trick}), the problem consists of computing
\begin{equation}\label{finalcomp}
A(\omega)=\frac{1}{2i\omega\beta(\omega)}\left[{\cal I}_1+{\cal I}_2
-\psi_0(2M)e^{-i\omega r^*_{\rm start}}+i\omega\int_{r^*_{\rm
start}}^{\infty}\psi_0(r)y_L(r^*,\omega)\,dr^*
\right]\ ,
\end{equation}
where ${\cal I}_1+{\cal I}_2$ is the integral given in (\ref{I1andI2})
and $\psi_0$ is given in (\ref{psi0}).  A numerical solution for
$A(\omega)$ is found by using a fourth-order Runge-Kutta routine to
solve for $y_L$ and $dy_L/dr^*$, and the integral in (\ref{finalcomp})
is done by Simpson's rule. From the solution for $A$ the energy
spectrum is computed with (\ref{spectrum}), and the waveform from
(\ref{inverse}). 

The numerical solution used a routine to find the ``particle
contribution,'' ${\cal I}_1+{\cal I}_2$, and one for the ``initial
value contribution,'' the integral over $\psi_0$ in (\ref{finalcomp}).
For both routines, second-order convergence was found and Richardson
extrapolation was used. The step size in the Runge-Kutta and
integration routines were halved until the Richardson extrapolate
agreed with that from the next larger grid within a preset error
limit. The initial value contribution could be
usually be found within an error of 0.2\%, while the particle
contributions required an error preset of 0.5\%.  For most values of
$\ell,\omega,r_0$, these precision requirements were easily met.  The
exception was a relatively small number of points at which the real
and imaginary parts of a contribution differed by more than an order
of magnitude. In this case it was difficult to get high accuracy in
the smaller part.

An estimate of the error in our results is complicated by the fact
that the physically important results are a superposition of the
particle and initial value contributions, and significant
cancellations occur in this superposition. These cancellations, in
principle, mean that the error may be much larger than the small
relative error in each contribution. To arrive at an estimate of the
error in our determinations of the radiated energy we have recomputed
the energy for four trials, in which $\pm0.5\%$ was added to the
particle contribution, and then to the initial value contribution.
The results for energies and errors are given in Table I. We see that
for many cases the computed energy is not highly sensitive to the
cancellation; the estimated error of around 1\% is just what we would
expect in the square (energy) of a quantity (amplitude) with an error
of 0.5\%. For some cases, however, especially those with higher
$\ell$, there is a significant magnification of error. We emphasize
that the error estimates given in Table I, are extremely
conservative. In arriving at them we have used the maximum 0.5\% error
applied to all values of $\omega$, whereas this maximum error actually
applies only to a small subset of the points. The smoothness of (most)
waveforms reported in the next section, and the consistent variation
of results with changing $r_0$, is evidence that the actual errors are
rather smaller than those reported in Table I.

In addition to the infall from finite $r_0$ we also have computed
spectra and waveforms for infall from infinity. We characterize these
cases with the same  parameter $\varepsilon_0$ we use for
infall from $r_0$. Here it has the value of the Lorentz $\gamma$-factor
\begin{equation}
\varepsilon_0\equiv1/\sqrt{1-v_\infty^2}
\end{equation}
for a particle with velocity $v_\infty$ at infinity.  For computation
of infall from infinity the above computational scheme is modified
only in the following ways: (i) The form of $T(r)$ in (\ref{time})
must be changed to 
\begin{eqnarray}
T(r)&=&-{ \varepsilon_0\over \varepsilon_0^2-1}\left({r\over2M}\right)
\sqrt{ \varepsilon_0^2-1+{2M\over r}}\nonumber\\
&&-{(2 \varepsilon_0^2-3) \varepsilon_0\over( \varepsilon_0^2-1)^{3/2}}
\ln\left[\sqrt{( \varepsilon_0^2-1)\left({r\over2M}\right)}+
\sqrt{1+( \varepsilon_0^2-1)\left(
{r\over2M}
\right)}\right]\nonumber\\
&&+\ln\left[{(2 \varepsilon_0^2-1)\left(
{r/2M}
\right)
+1+2 \varepsilon_0\left(
{r/2M}
\right)
\sqrt{ \varepsilon_0^2-1+
{2M/ r}
}\over\left(
{r/2M}
\right)-1}\right]\ .
\end{eqnarray}
(ii) The initial value contributions in (\ref{finalcomp}) must be
omitted. (iii) The limit of integration in (\ref{I1andI2}) must be
changed from $r_0^*$ to $\infty$. The computed energy for these
$ \varepsilon_0\geq1$ cases are given in Table II. Since there is only a
particle contribution in these cases, there is no issue of
cancellation affecting the errors.  The errors in the energy are, in
fact, primarily due to the cutoff in the solution at a finite radius
(with an analytic addition to represent the source contribution to
infinity). Error estimates were made by varying the cutoff radius, and
were found to be around 1\% for all $ \varepsilon_0\geq1$ cases.

\section{Waveforms and spectra}

Results for quadrupole waveforms and spectra naturally divide
themselves into three ranges, small $r_0/2M$ (less than $\sim2$),
moderate $r_0/2M$ (from $\sim2$ to $\sim5$) and large $r_0/2M$.
Waveforms are given as functions of retarded time $u\equiv t-r^*$.
For small $r_0$, as shown in Fig.\ \ref{smallr0}a, the shape of the
waveform is that of simple quasinormal ringing. This shape is the same
for all $r_0$ and the single example shown suffices for all small
$r_0$. Since the waveforms have the same shape, the energy spectra,
shown in Fig.\ \ref{smallr0}b,c, also have the same shape, changing
only in magnitude as $r_0$ increases.

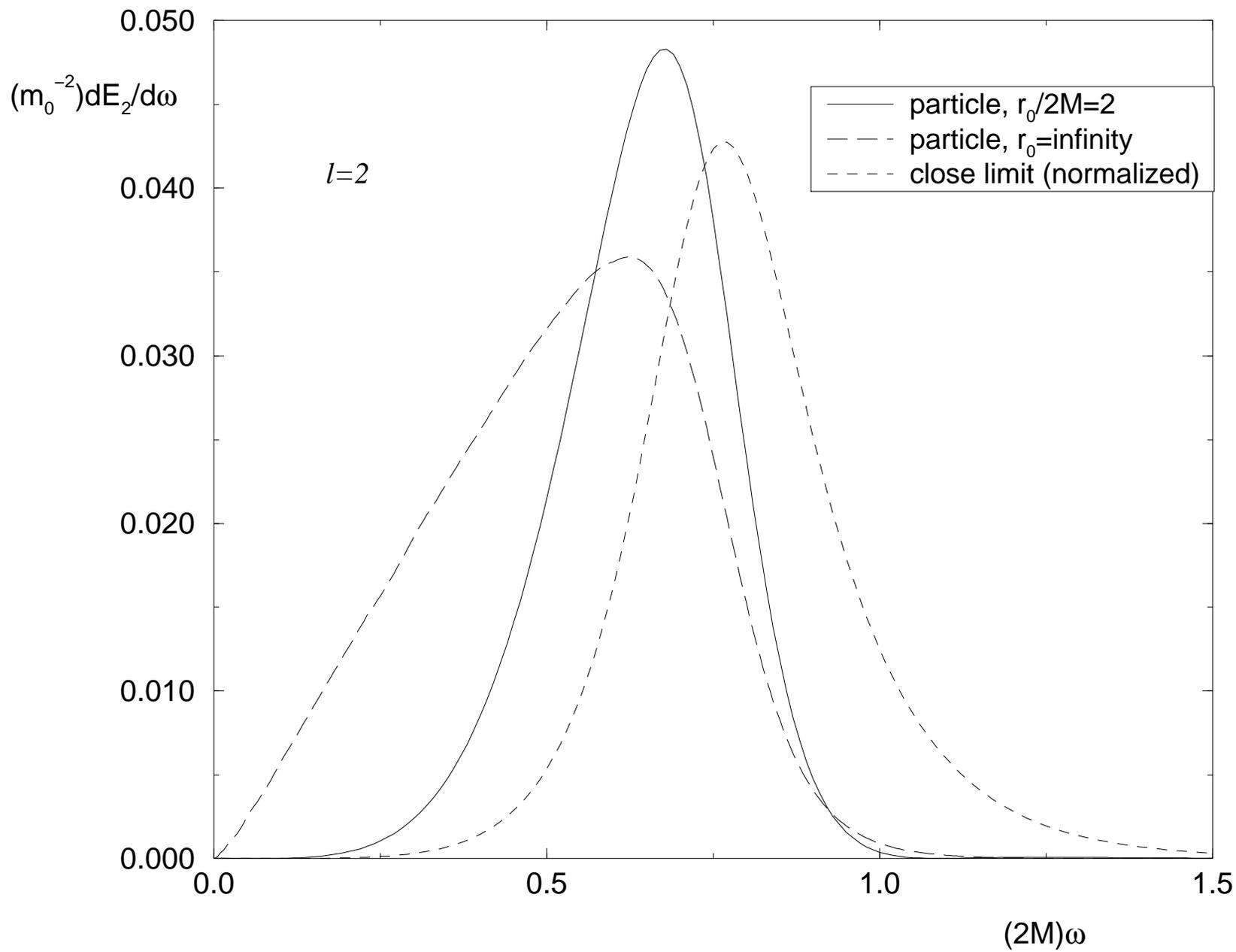
\begin{figure}
%fig 2
%\epsfxsize=200pt \epsfbox{fig1.eps}
%\epsfxsize=200pt \epsfbox{fig2.eps}
\caption{The $\ell=2$ spectra for $r_0/2M=2$,
and for the two limiting cases:
DRPP and close-limit. The close-limit case is normalized to have the
same energy as the $r_0/2M=2$ spectrum.  For the three spectra,
close-limit, $r_0/2M=2$, and DRPP, the maximum of the spectra occur
respectively at $\omega_{\rm max}(2M)=$ 0.765, 0.675,
and 0.625.}\label{specshapes}
\end{figure}

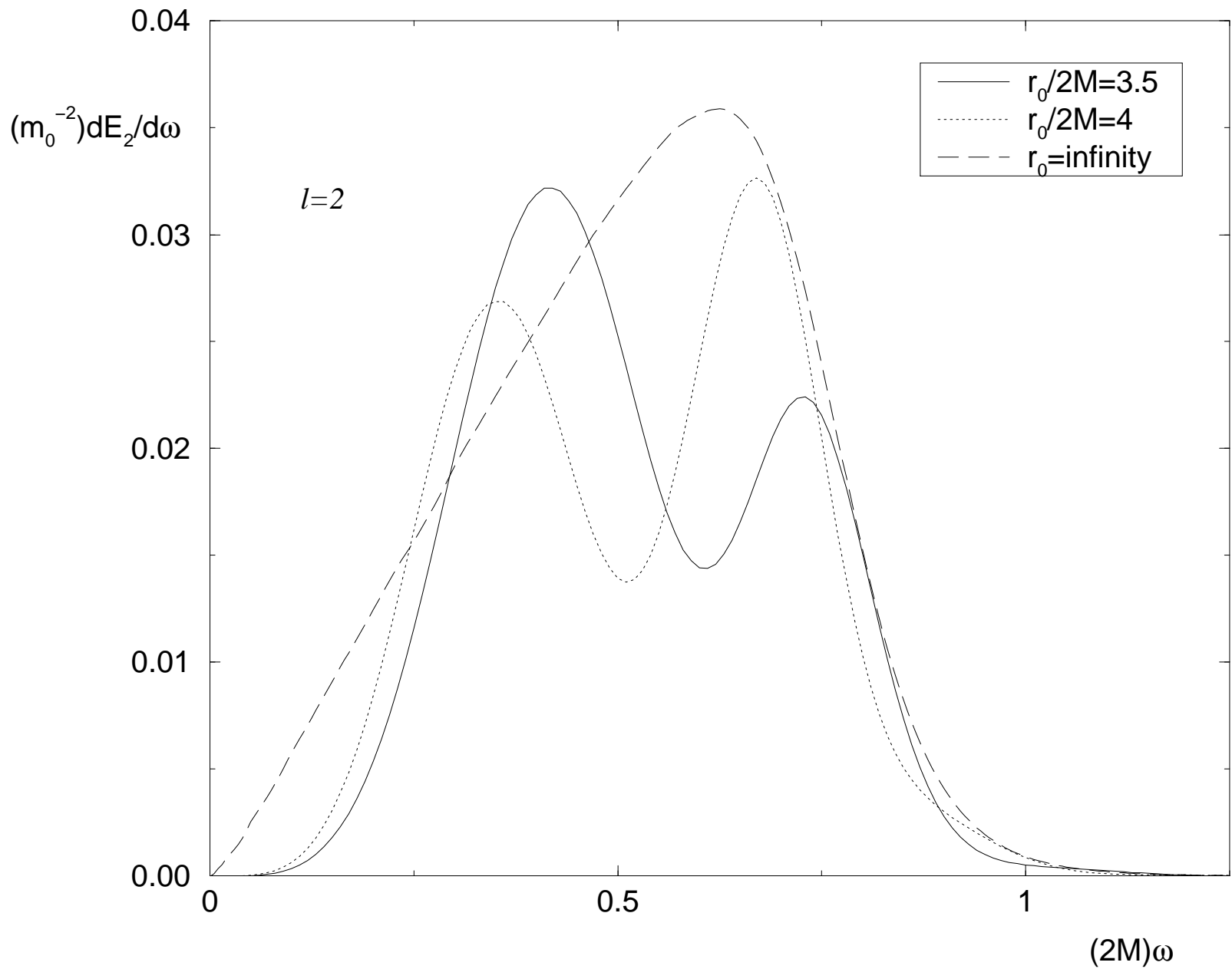
\begin{figure}
%fig 3
\caption{Spectra of $\ell=2$ energy for $r_0=3.5$ and 4. For comparison,
the DRPP spectrum for infall from infinity is also
shown.}\label{medr0}
\end{figure}

As $r_0$ increases further, the early negative excursion of the
waveform begins to broaden, and the spectrum shifts slightly.  Fig.\ 
\ref{specshapes} shows the $r_0/2M=2$ spectrum along with the spectra
for the two limiting cases, the DRPP infall from rest at $r_0
\rightarrow\infty$, and the close-limit spectrum $r_0 \rightarrow0$.
(The latter is normalized to have the same energy as the $r_0=2$
spectrum.) The $r_0/2M=2$ spectrum has an appearance that interpolates
between the two limits, as might be expected. As $r_0$ increases
further, however, changes in the spectrum develop that might not be
expected.  As shown in Fig.\ \ref{medr0}, the simple spectrum for
small $r_0$ develops a secondary peak and the secondary peak grows
with $r_0$.  As $r_0$ continues to increase, the initial shape of the
waveform becomes a very broad depression extending from the moment
infall begins to the beginning of quasinormal ringing. This is
illustrated in the waveform in Fig.\ \ref{hir0}a, for $r_0/2M=10$. The
start of infall, at $t=0, r_0/2M=10$ corresponds to $u/2M=-12.2$, and
it is at this value of $u$ that the waveform begins to take on nonzero
values. The initial part of the waveform, then, represents the
gravitational bremsstrahlung from the early nonrelativistic part of
the particle motion. The small wiggles around $u/2M\approx-12$ are a
numerical artifact due to imperfect cancellation of contributions from
the initial value and particle parts of the source. (To verify this we
changed the initial value contribution by $\pm10\%$ and found the
change in the initial wiggles to be much greater than in other
features of the waveform). For comparison, a DRPP waveform, for
infall from infinity, is also shown in Fig.\ \ref{hir0}.
(For infall from
infinity, of course, the zero of time cannot be set to the beginning of
infall. The time was arbitrarily shifted for the DRPP curve.) This
waveform has a similar ringing pattern as the infall from finite
radius, but lacks the initial waveform depression.

The end of the initial waveform depression is, roughly, the beginning
of of quasinormal ringing, as can be seen in Fig.\ \ref{hir0}a. The
generation of quasinormal ringing\cite{cmpodd} is associated with the
peak of the potential in (\ref{zpotential}). The time $t/2M\approx54$
at which the particle reaches the peak, at around $r/2M\approx1.5$ is
at retarded time $u/2M\approx53$. This is consistent with Fig.\
\ref{hir0}a, which shows ringing beginning somewhere around this value
of $u$.

It is interesting to compare the waveforms for a particle falling from
rest, to the waveform of a particle on a time-symmetric geodesic
trajectory, a particle that long in the past was moving radially
outward just outside a hole, that reaches a certain maximum radius
$r_0$ at time $t=0$, and that subsequently falls into the hole. The
analysis of this case requires only a simple modification of
(\ref{sofar}); the initial value terms are omitted and (due to time
symmetry of the source) the complex conjugate of ${\cal I}_1+{\cal
  I}_2$ is added to ${\cal I}_1+{\cal I}_2$. The resulting waveform is
shown in Fig.\ \ref{Poisson}a, and has also been given in Ref.\ 
\cite{annetal}. That waveform shows two periods of quasinormal ringing,
an early one excited when the particle goes outward through the region
around the potential peak, and a later one due to motion inward
through the peak. This later period of ringing, and in fact all
features of the waveform generated after $u/2M\approx-10$ agree very
closely with the waveform for the particle falling from rest. Because
the earlier ringing has a higher amplitude, the energy spectrum for
the symmetric trajectory, shown in Fig.\ \ref{Poisson}b, is dominated
by this early ringing, and has very large amplitude.  In the figure it
is seen to be much larger than the DRPP spectrum, which (see Fig.\ 
\ref{hir0}) approximates the spectrum for infall from $r_0/2M=7.5$.

Since the spectra for infall from both small $r_0$ and from $\infty$
have a single peak (see, e.g., Fig.\ \ref{specshapes}) it is
interesting that for intermediate values of $r_0$ the spectra are
characterized by a row of more-or-less evenly spaced bumps. As $r_0$
grows, the number of bumps increases, and the bumps decrease in
spacing and in height.  The origin of these modulations can be
understood by considering a simple example: Suppose a waveform
consists of nothing but a single period of quasinormal ringing with
the transform $A(\omega)$. The transform of the same waveform shifted
later in time by $T_{\rm shift}$ would be $A(\omega)e^{i\omega T_{\rm
    shift}}$. A wave consisting of {\em two} periods of ringing, the
original and the later one, would then have a transform
$A(\omega)(1+e^{i\omega T_{\rm shift}})$, and hence an energy spectrum
\begin{equation}
\omega^2|A(\omega)|^24\cos^2\left(\textstyle{\frac{1}{2}}\omega
T_{\rm shift}\right)\ .
\end{equation}
The combined spectrum would have the shape of the single waveform
spectrum modulated on a frequency scale $\delta\omega=2\pi/T_{\rm
shift}$. If the two waveforms were not identical, we would expect
modulation of the spectrum, but less than 100\% modulation.

This explanation can be tested on time-symmetric motion, where there
are two well separated periods of ringing in in Fig.\ \ref{Poisson}a.
The amplitudes of the ringing are different, of course, so we
shouldn't expect 100\% modulation of the spectrum and, indeed, the
modulations in Fig.\ \ref{Poisson}b, are not 100\%.  The time shift
between the first and second ringing periods is on the order of
$75(2M)$ (roughly the time it takes for the particle to rise up from
$r_0/2M=1.5$ to $r_0/2M=7.5$ and fall back to $r_0/2M=1.5$). This
suggests that the spacing of the spectral bumps should be
$\delta\omega=2\pi/[75(2M)]=0.84/2M$, which is in good agreement with
what is seen in Fig.\ \ref{Poisson}b. The application to a
nonsymmetric waveform, like that in Fig.\ \ref{hir0}a, is less
obvious.  The spacing of bumps $\delta\omega\approx0.11/2M$ in Fig.\ 
\ref{hir0}c suggests that the time shift is on the order of $57(2M)$.
This, presumably, represents the time between the descent of the
waveform at $u/2M\approx0$ and the start of ringing at
$u/2M\approx50$.  

The interpretation works as well as a predictor of the bump spacing
for other values of $r_0$, and seems clearly to be qualitatively
correct, and to explain the progression of the spectra.  As $r_0$
approaches the horizon, the time between the initial moment and the
onset of ringing goes to zero, so the spacing of bumps is infinite and
there are no modulations of the spectrum. We get the single-humped
close-limit spectrum. As $r_0$ grows, the time between the initial
motion and the start of ringing increases, so the spacing of the bumps
gets smaller, and hence more bumps appear. As $r_0$ is becoming
larger, however, the early waveform is becoming less dramatic (the
initial depression is decreasing in amplitude) so in interacting with
the later ringing it is producing smaller modulation; the height of
the bumps is decreasing. Finally, as $r_0\rightarrow\infty$, we can
think of the spectrum approaching one that has an infinite number of
infinitesimally spaced zero-height bumps, bumps that are invisible in
the DRPP spectrum.

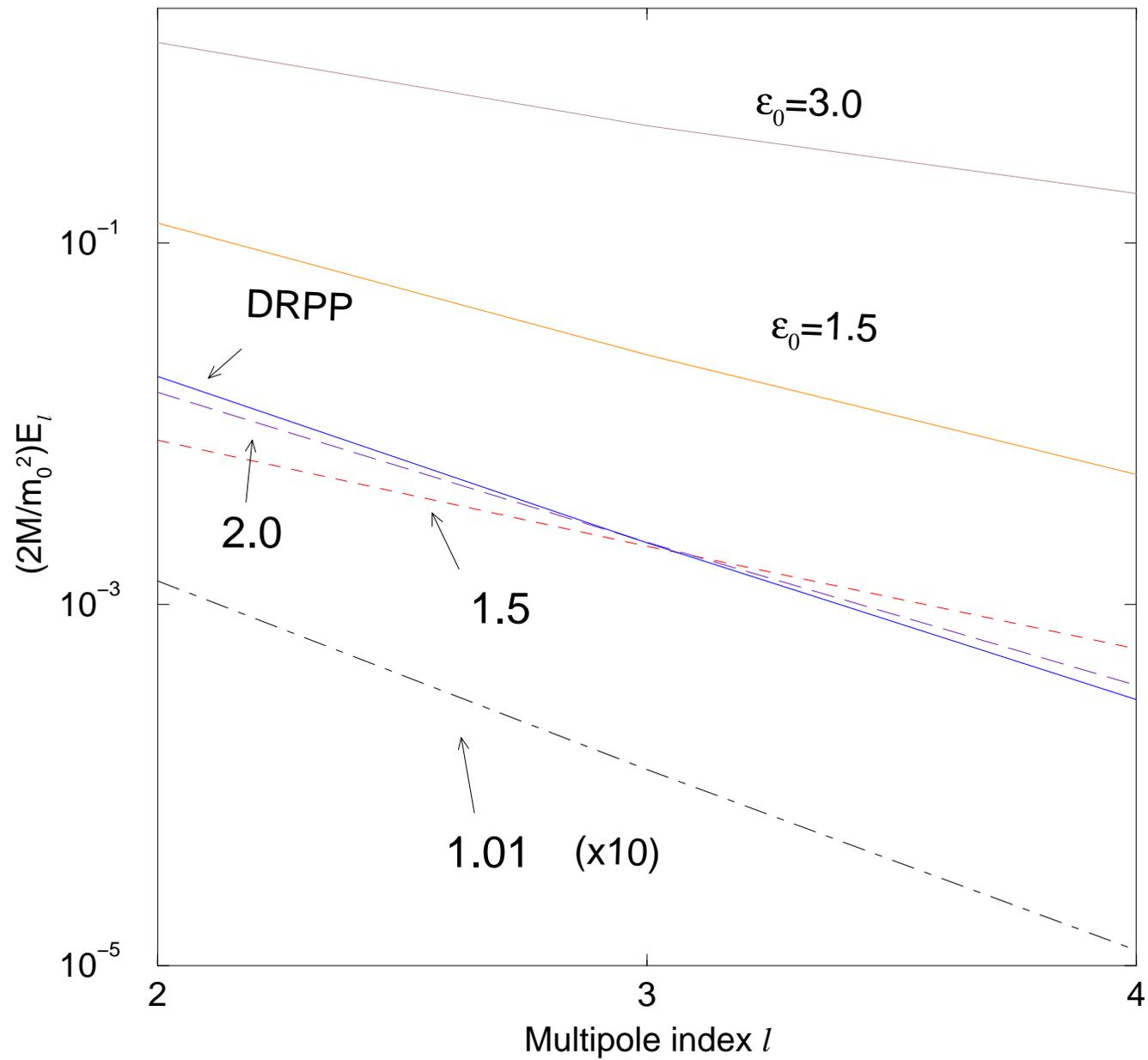
\begin{figure}
%fig 7
%\epsfxsize=200pt \epsfbox{fig1.eps}
%\epsfxsize=200pt \epsfbox{fig2.eps}
\caption{Energy in different multipoles. Energy for $\ell=2,3,4$ is shown
for several values of $r_0$ in the case of infall from finite radius, and
for several different values of $\varepsilon_0$ in the case of infall from
infinity.  The energies for $r_0/2M=1.01$ are multiplied by 10 to
improve the plot.}\label{evl}
\end{figure}

Results for higher $\ell$-poles are shown in Figs.\
\ref{diffells}. The waveforms for $\ell>2$ show a more complicated
structure of the pre-ringing radiation, resulting in more complex
spectra. The total energy radiated in different $\ell$-poles is shown
in Fig.\
\ref{evl}. Results are given for several different values of $r_0$,
and for particles falling in from infinity, both with no initial
velocity (the DRPP case) and with $ \varepsilon_0>1$. The distribution of
energy among the multipoles is dominated by the quadrupole for small
$r_0$. As $r_0/2M$ increases to around 1.5, the higher multipoles
become more important, but with further increase of $r_0/2M$ to
$\sim2$ the trend is reversed, and the ratio of multipole energies
take on values that remain constant for further increases in $r_0$.
For infall from infinity, the ultrarelativistic cases, with high
$ \varepsilon_0$, radiate more heavily in higher multipoles, as would be
expected.  To find the total energy, given in Fig.\ \ref{evdist}, we
computed energy radiated in $\ell=2,3,4$ modes and assumed that the
energies from each $\ell$ decreased as a geometric series.  This
allowed us to add an estimate of the contributions from $\ell>4$. The
addition was typically around 2\% of the energy. The energy is plotted
as a function of proper distance $L=\int_{2M}^{r_0}dr/\sqrt{1-2M/r}$,
rather than $r_0$, to show more clearly the details for small
separation.

Aspects of infall from infinity are given in Fig.\ \ref{gammas}, which
shows waveforms and spectra.  The waveforms are characterized by a
large early amplitude that is an increasing function of $
\varepsilon_0$, and the spectra show strong low frequency radiation,
extending to $\omega=0$, due to this early phase of the radiation. The
total energy radiated in the lowest three multipoles is also
shown. (Since the relative importance of higher multipoles increases
with $ \varepsilon_0$, extrapolation to total radiated energy is not
immediate.) The energy results are extended to include infall from
finite radius where $\varepsilon_0$ takes the value
$\sqrt{1-2M/r_0}$.

\section{Discussion}

One of the interesting questions that can be clarified with the above
results is the validity of the close approximation. This
approximation, for two holes, assumes that the holes are initially
close enough so that the structure of the initial data at small radius
is inside an initial all-encompassing horizon. Only the large $r$
features of the initial data therefore are relevant to the production
of outgoing radiation. We can immediately apply this method to the
particle problem by comparing our initial geometry, in
(\ref{ivmetric}) and (\ref{Fform}), with the initial geometry in
equation (4.25) of Ref.~\cite{annetal}. We see that the results of
that reference can be applied to the particle problem by the
replacement
\begin{equation}
8M\kappa_\ell(\mu_0)\rightarrow m_2(z_2/M)^\ell\ .
\end{equation}
With (\ref{z2eq}) and (\ref{massreln}) this can be rewritten as 
\begin{equation}\label{replace}
\kappa_\ell(\mu_0)\rightarrow \frac{1}{2^{\ell+2}}
\left(
\frac{m_0}{2M}
\right)
\left(
\frac{r_0}{2M}
\right)^\ell
\left(
1+\sqrt{1-2M/r_0
}
\right)^{2\ell+1}\ .
\end{equation}
By the methods of Refs.~\cite{annetal,pp}, $E_\ell/2M$, the radiated
energy in units of $2M$, for each multipole, is shown to be
$1.26\times10^{-2}\kappa_2^2, \ 3.10\times10^{-3}\kappa_3^2,
\ 8.33\times10^{-4}\kappa_4^2,\cdots$, respectively for
$\ell=2,3,4\cdots$. If we replace $\kappa_\ell$ with (\ref{replace})
we find the close limit predictions for energy.  In
Refs.~\cite{annetal,pp} the outgoing radiation is computed from an
evolution of $\psi_\ell$, using a finite difference representation of
(\ref{rtzerilli}) (with no source term). Here we have also computed
waveforms and energies in the close limit  directly
by the transform methods of Secs.~II and III, with only the following
changes: The integrals ${\cal I}_1+{\cal I}_2$ are omitted from
(\ref{sofar}), and only the $\bar r>z_2$ form of 
${\cal F}_\ell(\bar r)$ is used
in (\ref{Fform}). The results of the two methods are energy values
that agree to better than 1\% and waveforms that are almost
indistinguishable. (The numerically evolved waveforms are rather
smoother than the waveforms from the transform method. They lack the
small amplitude wiggles that can be seen, e.g., in the waveform in
Fig.\ \ref{smallr0}a at early times.)

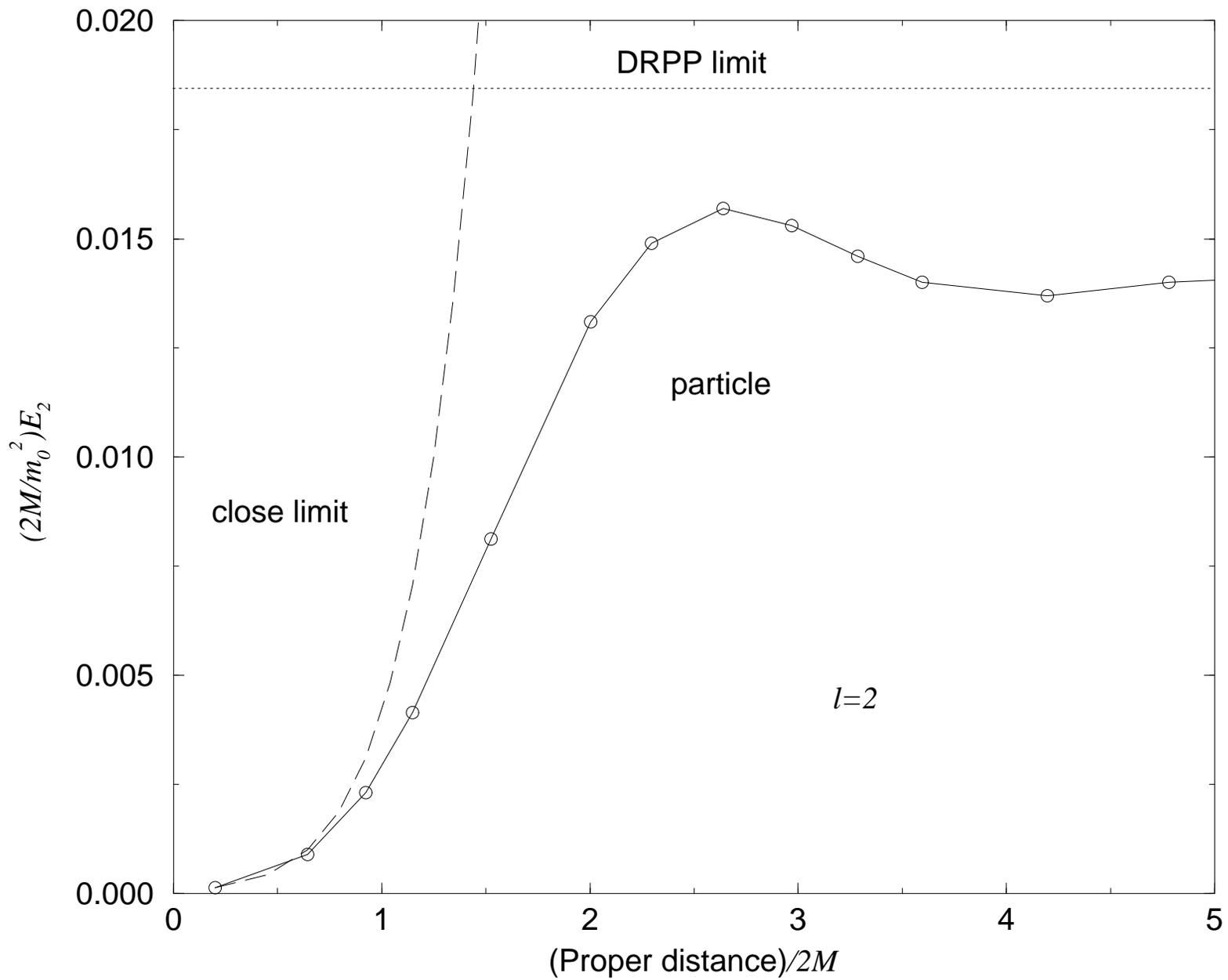
\begin{figure}
%fig 10
%\epsfxsize=200pt \epsfbox{fig1.eps}
%\epsfxsize=200pt \epsfbox{fig2.eps}
\caption{Quadrupole energy for an infalling particle as a function of 
the particle's initial proper distance from the horizon. The computed
energy is compared with the close-limit
approximation.}\label{closelim}
\end{figure}

In Fig.\ \ref{closelim}, we plot the $\ell=2$ close-limit energy
prediction and compare it to the full computation for the particle
infall. It is clear that the close limit method is acceptable out to
$r_0\approx 2.2M-2.3M$, and fails by a large factor at $r_0=3M$. This
is in accord with the general picture that the close limit should be a
reasonably good approximation when the particle starts inside the
peak, around $r\approx3M$, of the potential (\ref{zpotential}).  The
waveforms in the close limit all have precisely the same shape; only
the amplitudes differ. The computed waveforms are in excellent
agreement with the small $r_0$ waveforms (see, e.g., the waveforms for
$r_0/2M=1.5$ in Fig.\ \ref{smallr0}a). The close-limit predictions of
waveform shape are even better than the energy predictions. Only for
$r_0/2M$ larger than around 1.5 does the waveform start to change from
the close-limit shape.

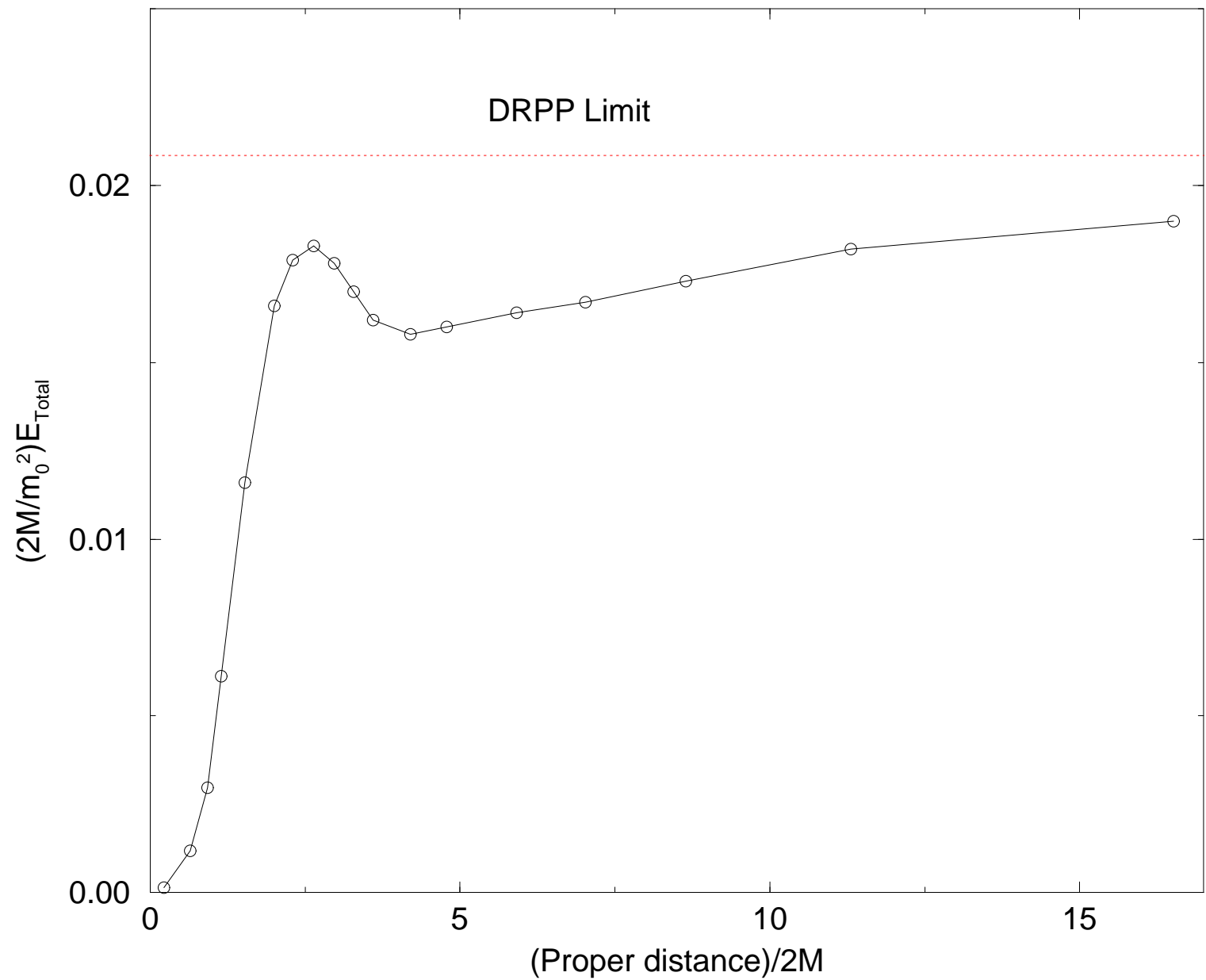
\begin{figure}
%fig 8
%\epsfxsize=200pt \epsfbox{fig1.eps}
%\epsfxsize=200pt \epsfbox{fig2.eps}
\caption{Total energy radiated by a falling particle, as a function
of the initial proper distance of the particle from the horizon. The
points shown are at $r_0/2M=$ 1.01, 1.1, 1.2, 1.3, 1.5, 1.8, 2, 2.25,
2.5, 2.75, 3, 3.5, 4, 5, 6, 7.5, 10, 15. The local maximum is at
$r_0/2M\approx2.25$ and the minimum at $r_0/2M\approx3.5$.
}\label{evdist}
\end{figure}

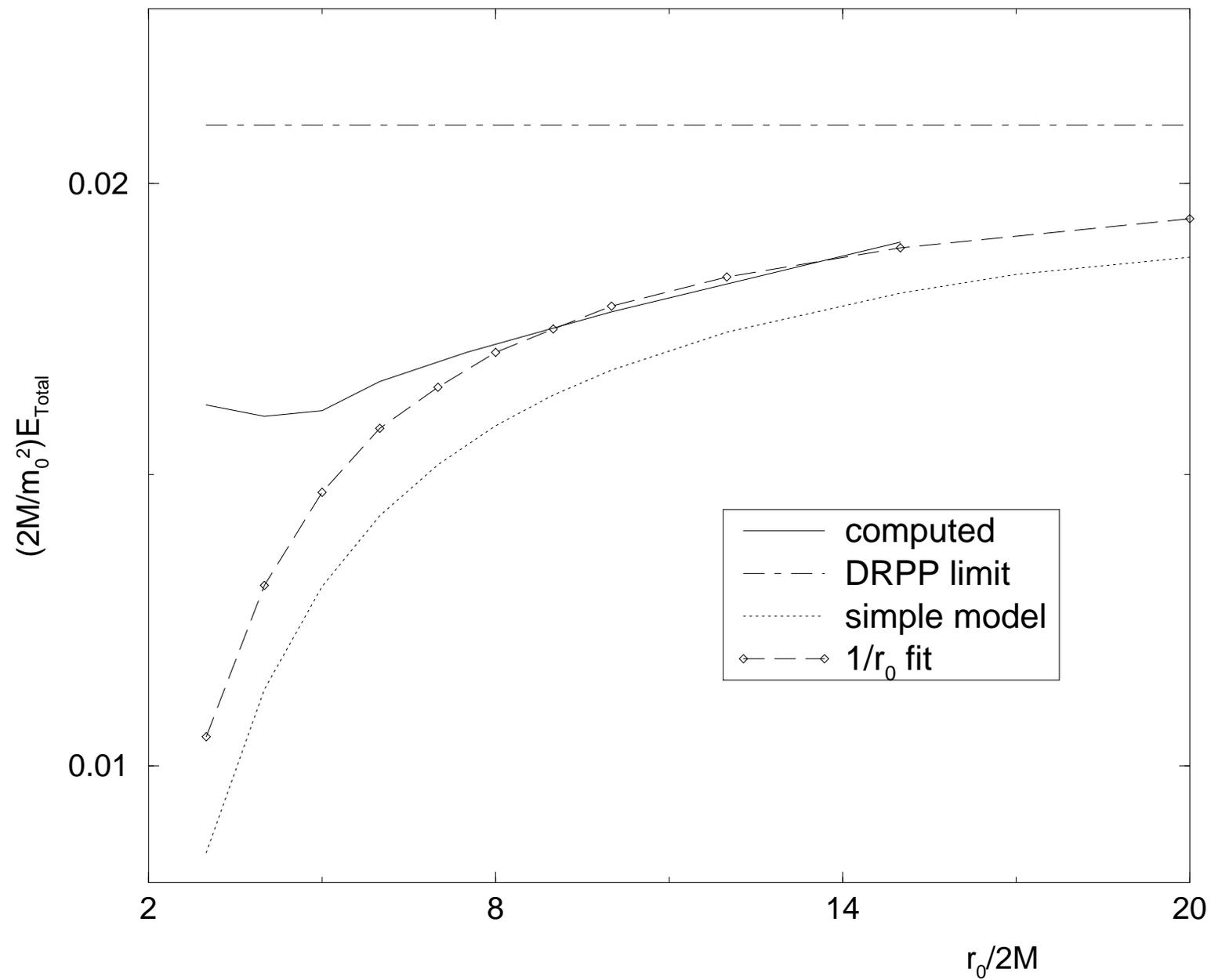
\begin{figure}
%fig 11
%\epsfxsize=200pt \epsfbox{fig1.eps}
%\epsfxsize=200pt \epsfbox{fig2.eps}
\caption{Total radiated energy as a function of $r_0/2M$. Computed results
are compared with a simple model, and with the best $1/r_0$
fit.}\label{farapprox}
\end{figure}

The dependence of radiated energy on $r_0$, shown in Fig.\
\ref{evdist}, is perhaps the most interesting result of our
computations.  One has an intuitive instinct that the radiated energy
should decrease with decreasing $r_0$. At the crudest level this
intuition is based on the idea that infall from a larger radius
results in a particle which ``strikes'' the black hole harder and
excites more quasinormal ringing. Since a decrease of energy with
smaller $r_0$ is expected, a natural first guess for approximating the
decrease is to multiply the DRPP energy by a reduction factor
$(1-k2M/r_0)$, where $k$ is some fitting parameter of order
unity. This approximation, with $k=1.5$ is shown in Fig.\
\ref{farapprox} and is compared with the computed results, and with
the DRPP limit. The choice $k=1.5$ was made to give good agreement at
$r_0/2M=15$, and presumably at larger values of $r_0$. (It is
difficult to compute energies at much larger values of $r_0$, due to
the rapid modulation of the spectrum.)

A more physical justification for decrease of radiation with
decreasing $r_0$ can be constructed starting with the quadrupole
formula. A faster moving particle from a larger $r_0$ implies a larger
value of the time derivatives of the quadrupole moment. This argument
has been used\cite{suenstuff} as the basis of a simple quantitative
model for the effect of varying $r_0$. The energy for infall from
infinity, in that model, is reduced by a reduction factor $F_{r_0}$
based on the quadrupole formula. [See eq. (23) of Ref.~
\cite{suenstuff}.] (For large $r_0$ that reduction factor reduces to
$F_{r_0}=1-(60/27)(2M/r_0)+{\cal O}(2M/r_0)^2$.)  In Fig.\ 
\ref{farapprox}, we show the result of that simple model. It is clear
that the $F_{r_0}$ factor captures the correct qualitative feature of
a decrease of radiation with decrease of $r_0$ but implies too
dramatic a decrease.

As $r_0$ continues to decrease, a rather unexpected effect appears.
Below $r_0\approx7M$ the energy begins to increase with decreasing
separation. At yet smaller radii ($r_0$ less than around 4.5M) the
energy again decreases with decreasing $r_0$, as the close limit
dictates it must.  Thus the relationship of radiated energy and $r_0$
has the expected nature in the two regimes where simple arguments
apply: large separations and small separations. The anomalous behavior
in the range $4.5M$--$7M$ underscores the fact that the generation of
outgoing radiation is tied closely to the nature of the potential
(\ref{zpotential}), which peaks around $r_0=3M$, and cannot be
understood in terms of close or far approximations.  This anomaly, it
should be noted, appears to have no equivalent feature in the case of
the head-on collision of two equal mass holes\cite{annetal}.
Presumably this is because the replacement of the particle by a hole
means that the infalling hole is not localized at a particular value
of the potential of the other hole. As the mass ratio of the infalling
holes becomes smaller and smaller there must come a point at which an
anomalous bump develops in the dependence of radiation on initial
separation.

We have seen that the particle limit provides a relatively easy tool
for understanding some aspects of the generation of radiation, and of
the collisions of holes. We intend next to use this formalism to study
what features on initial data are important to determining how much
energy is radiated for particle infall, and presumably for black hole
collisions.

\begin{acknowledgments}
This work has been partially supported by the National Science
Foundation under grant PHY0507719. We thank Eric Poisson for 
useful discussions of computational aspects of the problem.
\end{acknowledgments}

\pagebreak

\begin{figure}
%fig 1
%\epsfxsize=200pt \epsfbox{fig1.eps}
%\epsfxsize=200pt \epsfbox{fig2.eps}
\caption{
The $\ell=2$ waveforms and spectra for small $r_0$.  In (a) the
waveform is shown as a function of retarded time $u\equiv t-r^*$, with
$t=0$ corresponding to the moment at which the infall begins.
Included for comparison is the predictions of the close-limit
approximation.  For the $r_0/2M=1.1$ case, shown in (b), the
close-limit energy is larger than the computed energy by 11\%.  For
the $r_0/2M=1.5$ case, in (c), the close-limit energy is 2.9 times the
computed energy.}\label{smallr0}
\end{figure}

\begin{figure}
%figs 4
\caption{
The $\ell=2$ waveforms and spectra for large $r_0$. In (a) the waveform
 for $r_0/2M= 10$ is shown as a funtion of retarded time $u$, and is
 contrasted with the DRPP waveform (for which the zero of retarded
 time has a different meaning).  Energy spectra are shown in (b)--(d)
 for three large values of $r_0$ and are contrasted with the
 $r_0\rightarrow\infty$ DRPP limit (dashed curves).}\label{hir0}
\end{figure}

\begin{figure}
%fig 5
%\epsfxsize=200pt \epsfbox{fig1.eps}
%\epsfxsize=200pt \epsfbox{fig2.eps}
\caption{
The $\ell=2$ waveform and spectrum for a time-symmetric trajectory. In
 (a) the waveform (dotted curve) is given for a particle that moves
 outward and reaches a maximum at $r_0/2M=7.5$ before falling inward,
 and is compared with the waveform for infall from $r_0/2M=7.5$. In
 (b) the spectrum of energy generated by the time-symmetric motion is
 compared with the DRPP spectrum for infall from infinity. The DRPP
 spectrum contains less than 5\% as much energy as that for the
 time-symmetric motion.  }\label{Poisson}
\end{figure}

\begin{figure}
%fig 6
%\epsfxsize=200pt \epsfbox{fig1.eps}
%\epsfxsize=200pt \epsfbox{fig2.eps}
\caption{Results for higher multipole moments. For $r_0/2M=5$,
waveforms and spectra are shown
for $\ell=2,3,4$.  Dashed curves are spectra for infall from
infinity.}\label{diffells}
\end{figure}

\begin{figure}
%fig 9
%\epsfxsize=200pt \epsfbox{fig1.eps}
%\epsfxsize=200pt \epsfbox{fig2.eps}
\caption{Results for infall from infinity. Plots are given of waveforms
(a) and spectra (b), (c), for a particle falling in from infinity with
nonzero energy. The total radiation emitted in the first three
multipoles is shown as a function of $\varepsilon_0$. This curve also
extends to particles falling from finite radius, with
$\varepsilon_0=(1-2M/r_0)^{1/2}$.  The local maximum of the spectrum
occurs for $\varepsilon_0\approx0.75$, and the local minimum at about
$\varepsilon_0\approx0.84$.}
\label{gammas}
\end{figure}

\pagebreak

\begin{table}
\caption{Radiated energy for infall from $r_0$}
\bigskip
\begin{tabular}{lccc}
$r_0/2M$&${\ell}$&$(2M/m_0^2)E_{\ell}$&error\\
\tableline
\\
15&2&$1.64\times 10^{-2}$&1\%\\
\ &3&$1.98\times 10^{-3}$&1\%\\
\ &4&$2.88\times 10^{-4}$&5\%\\
\\
5&2&$1.43\times 10^{-2}$&1\%\\
\ &3&$1.62\times 10^{-3}$&2\%\\
\ &4&$2.23\times 10^{-4}$&6\%\\
\\
3&2&$1.40\times 10^{-2}$&2\%\\
\ &3&$1.65\times 10^{-3}$&4\%\\
\ &4&$2.61\times 10^{-4}$&11\%\\
\\
2&2&$1.49\times 10^{-2}$&3\%\\
\ &3&$2.21\times 10^{-3}$&5\%\\
\ &4&$3.56\times 10^{-4}$&11\%\\
\\
1.5&2&$8.11\times 10^{-3}$&3\%\\
\ &3&$2.10\times 10^{-3}$&5\%\\
\ &4&$5.66\times 10^{-4}$&10\%\\
\\
1.1&2&$9.02\times 10^{-4}$&$<1\%$\\
\ &3&$1.85\times 10^{-4}$&1\%\\
\ &4&$4.06\times 10^{-5}$&4\%\\
\end{tabular}
\label{t1}
\end{table}

\begin{table}
\caption{Radiated energy for infall from infinity}
\bigskip
\begin{tabular}{lccc}
$\varepsilon_0$&$(2M/m_0^2)E_2$&$(2M/m_0^2)E_3$&$(2M/m_0^2)E_4$\\
\tableline
\\
1&$1.82\times 10^{-2}$&$2.18\times 10^{-3}$&$2.96\times 10^{-4}$\\
1.1&$2.75\times 10^{-2}$&$3.48\times 10^{-3}$&$5.06\times 10^{-4}$\\
1.3&$6.48\times 10^{-2}$&$9.90\times 10^{-3}$&$1.73\times 10^{-3}$\\
1.5&$1.285\times 10^{-1}$&$2.40\times 10^{-2}$&$5.21\times 10^{-3}$\\
1.8&$2.70\times 10^{-1}$&$6.34\times 10^{-2}$&$1.76\times 10^{-2}$\\
3&$1.285$&$4.463\times 10^{-1}$&$1.885\times 10^{-1}$\\
\end{tabular}
\label{t2}
\end{table}

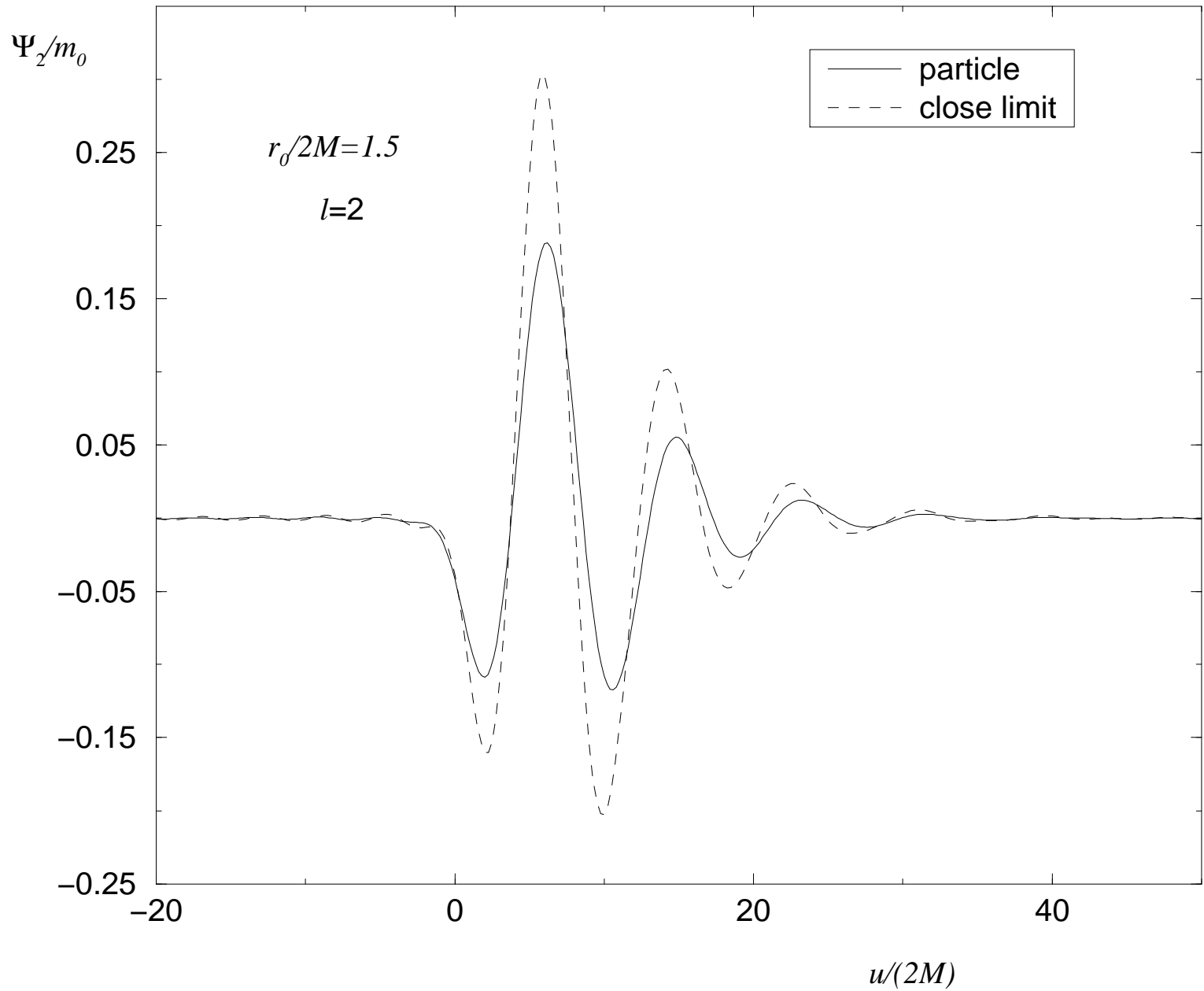
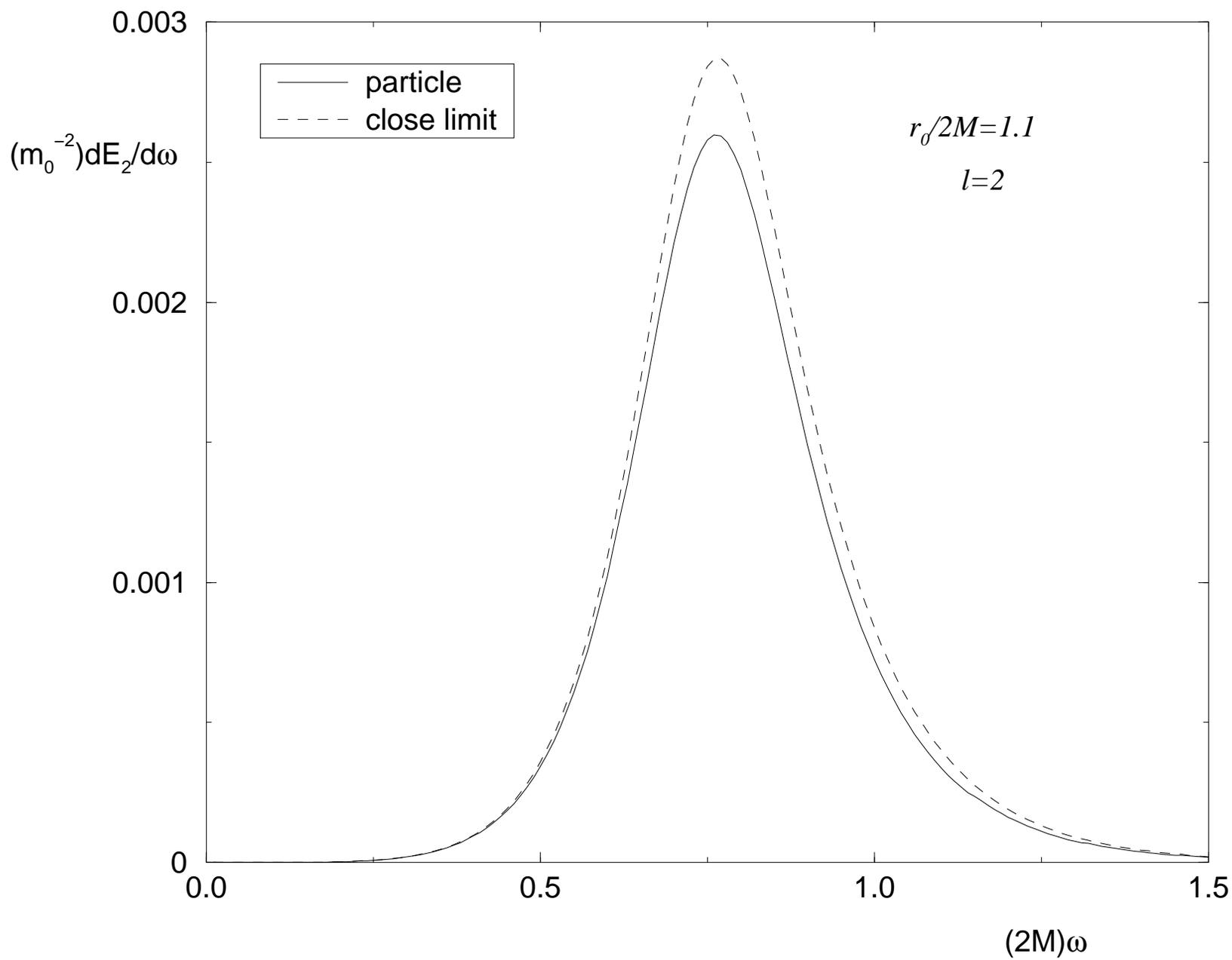
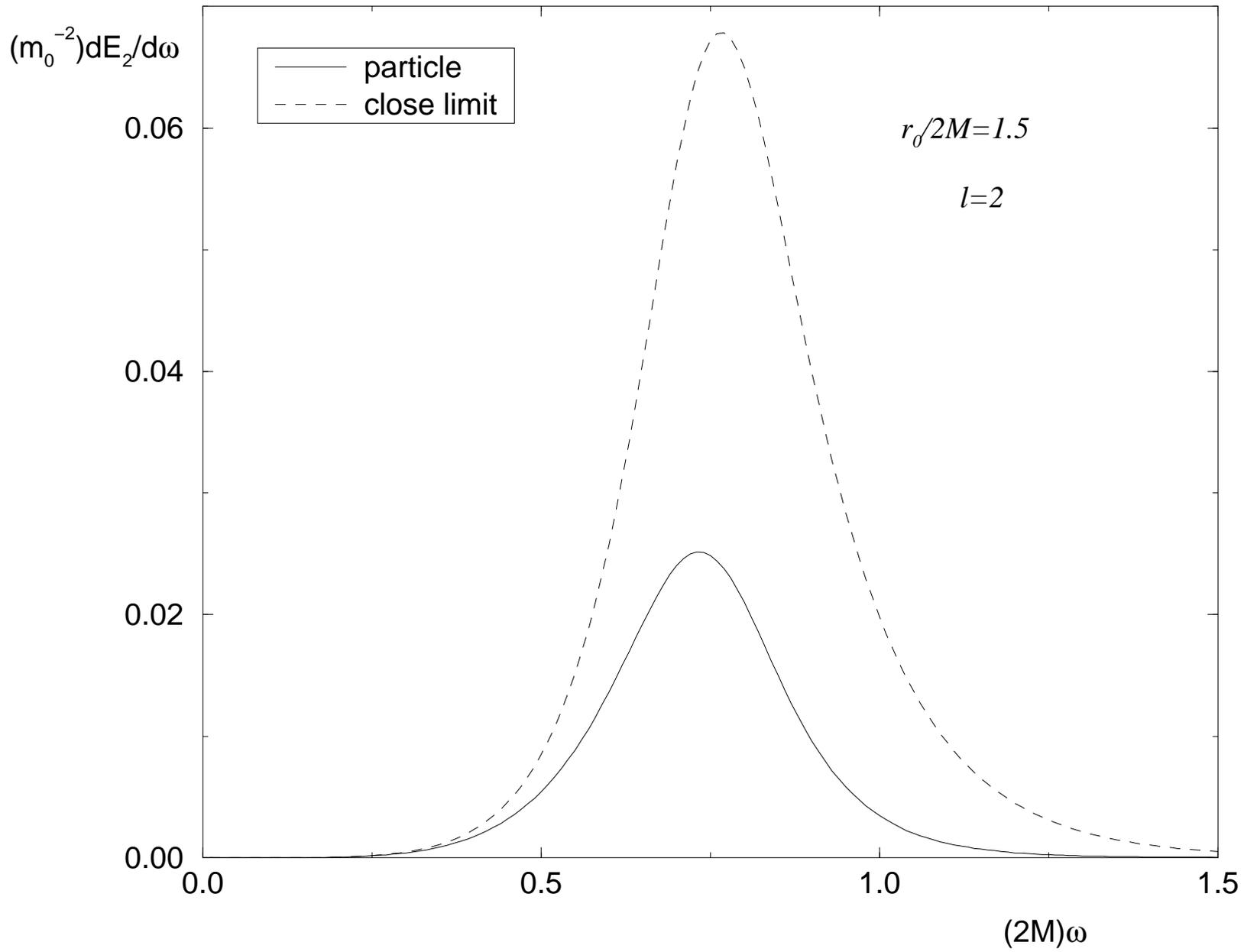
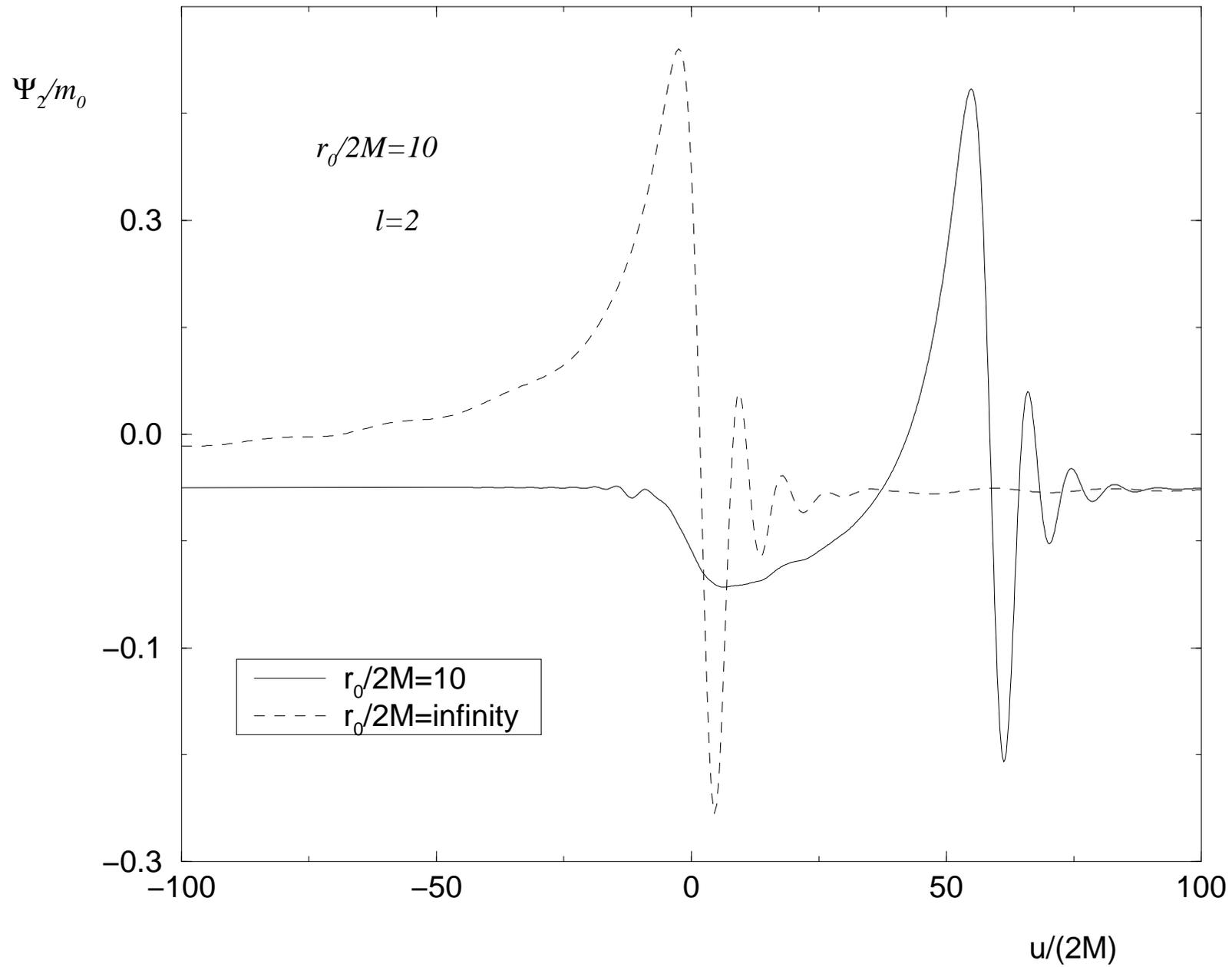
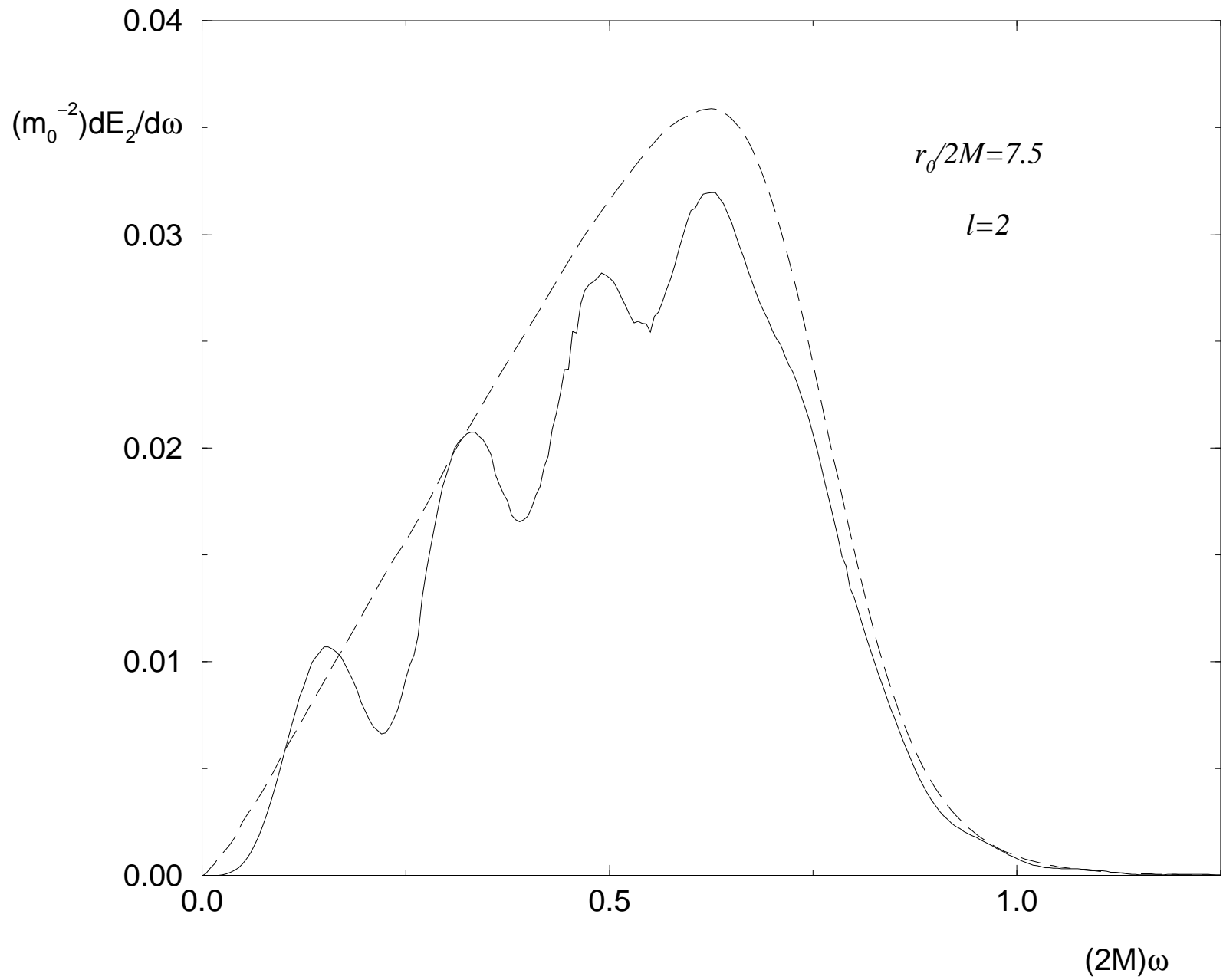
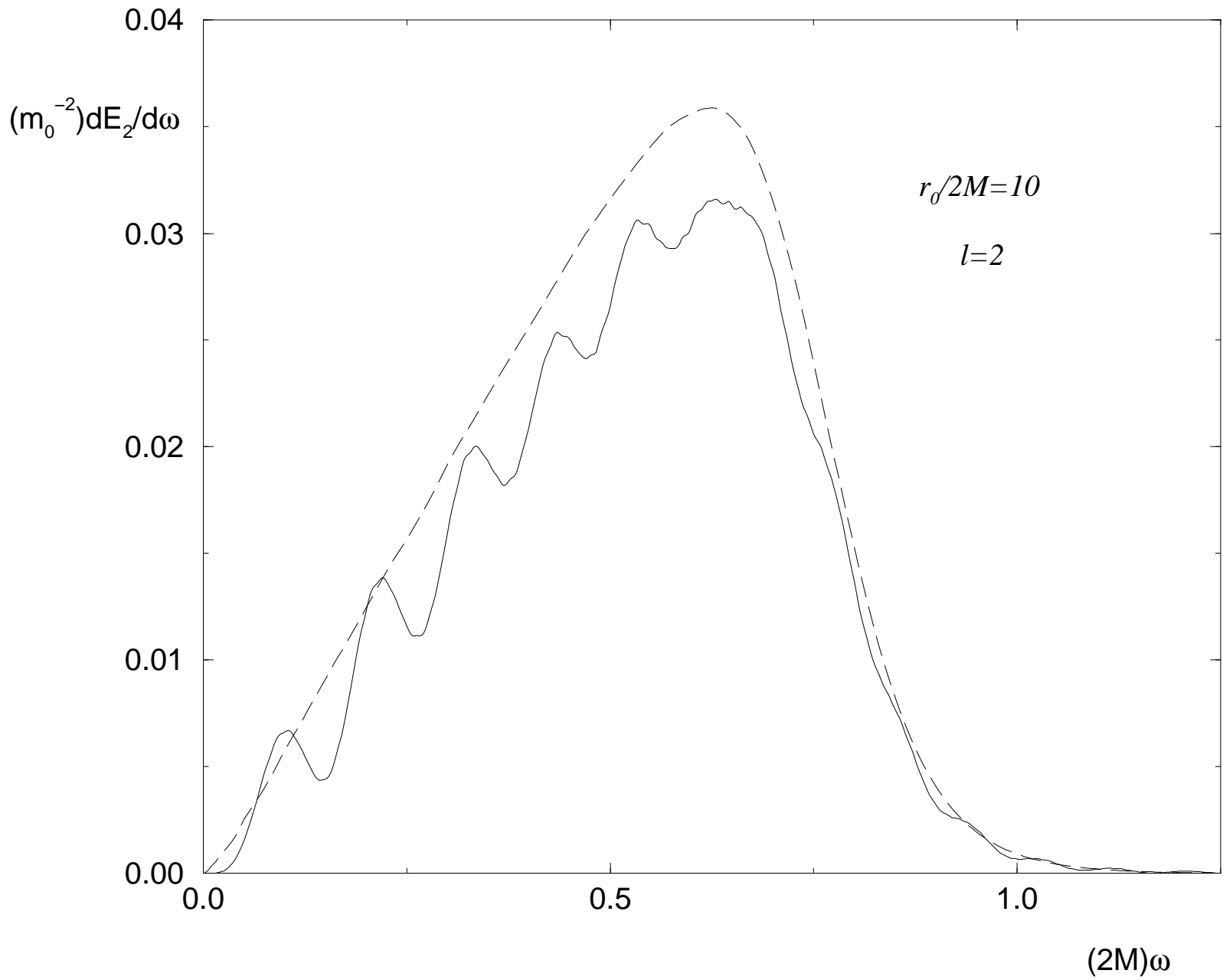
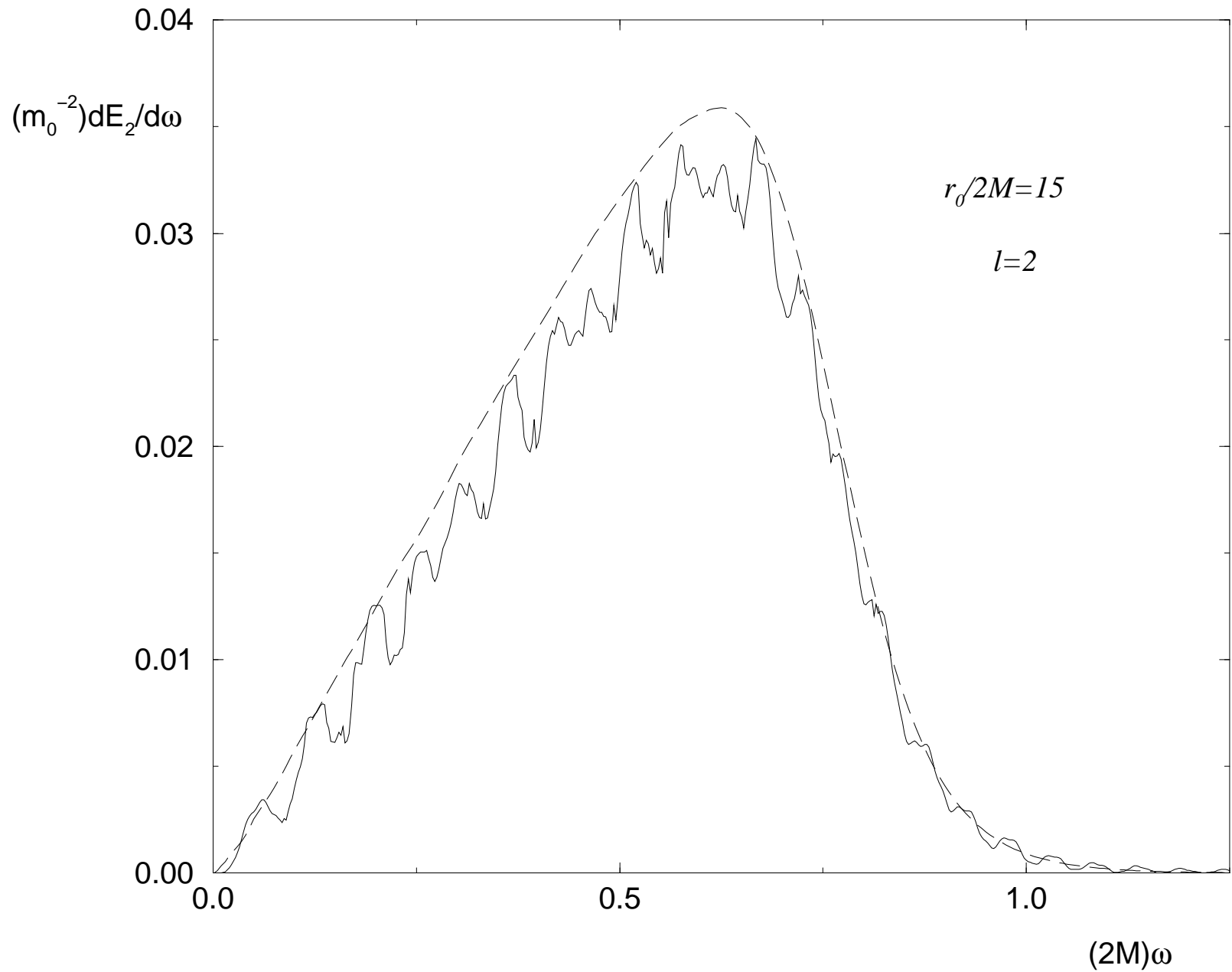
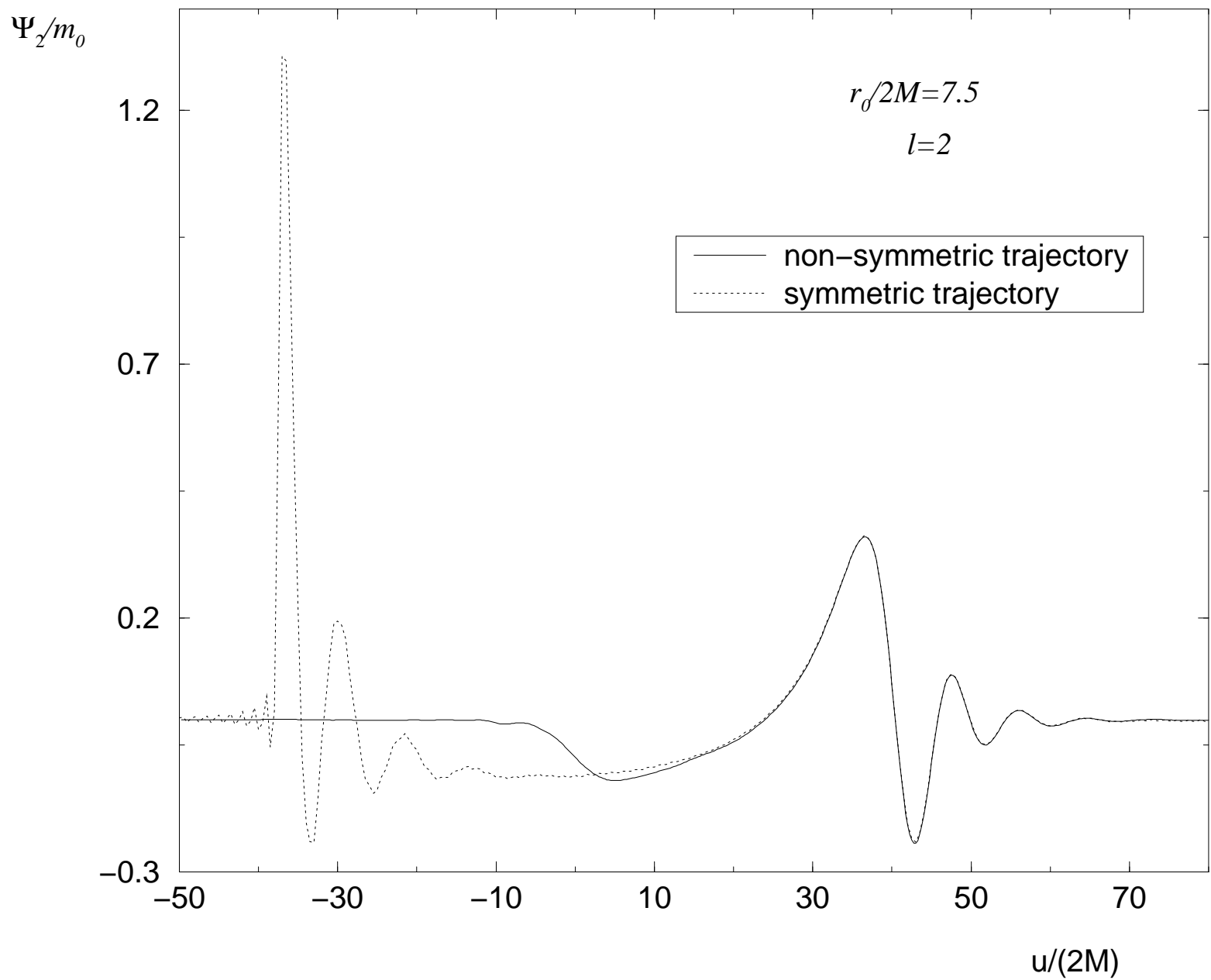
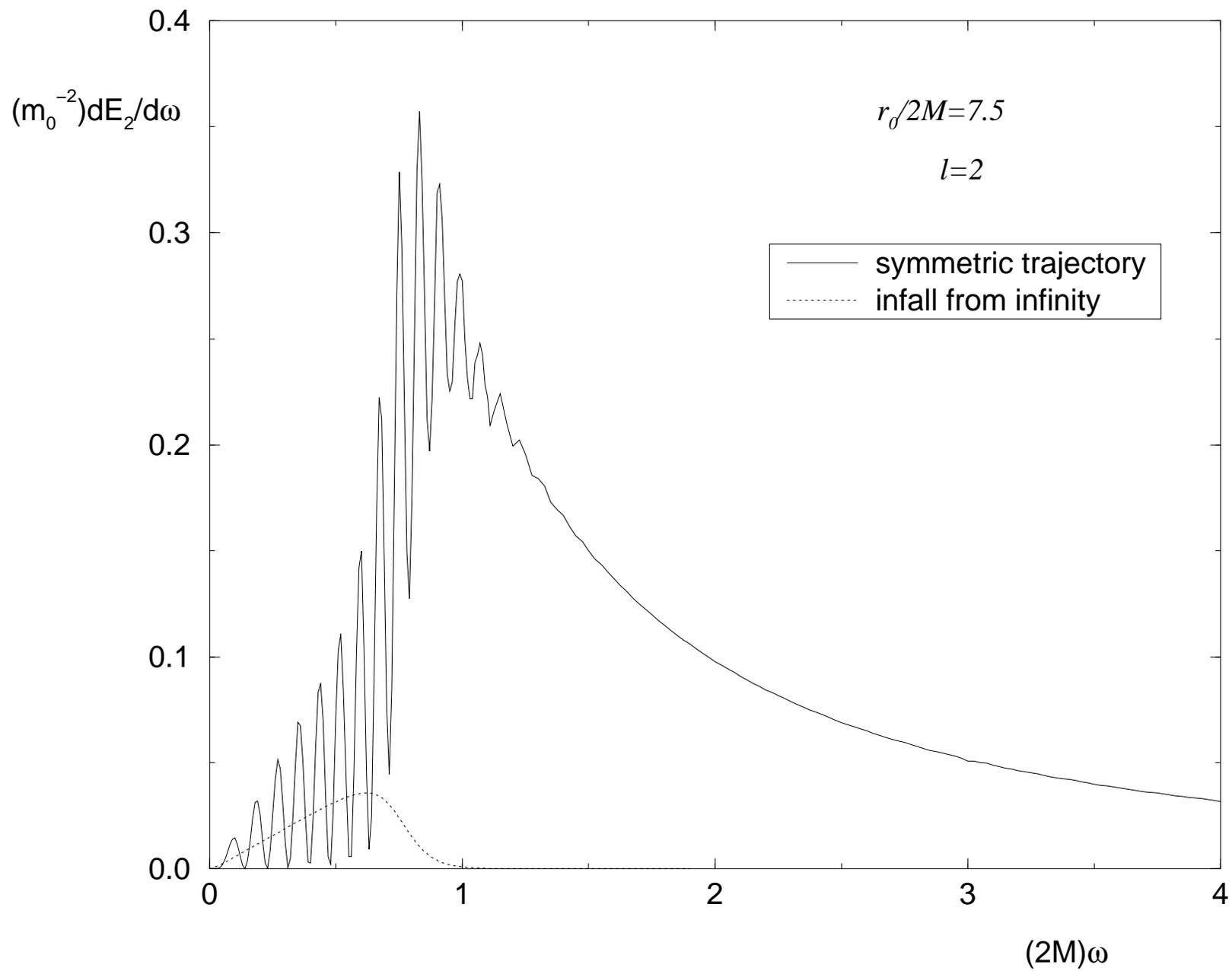
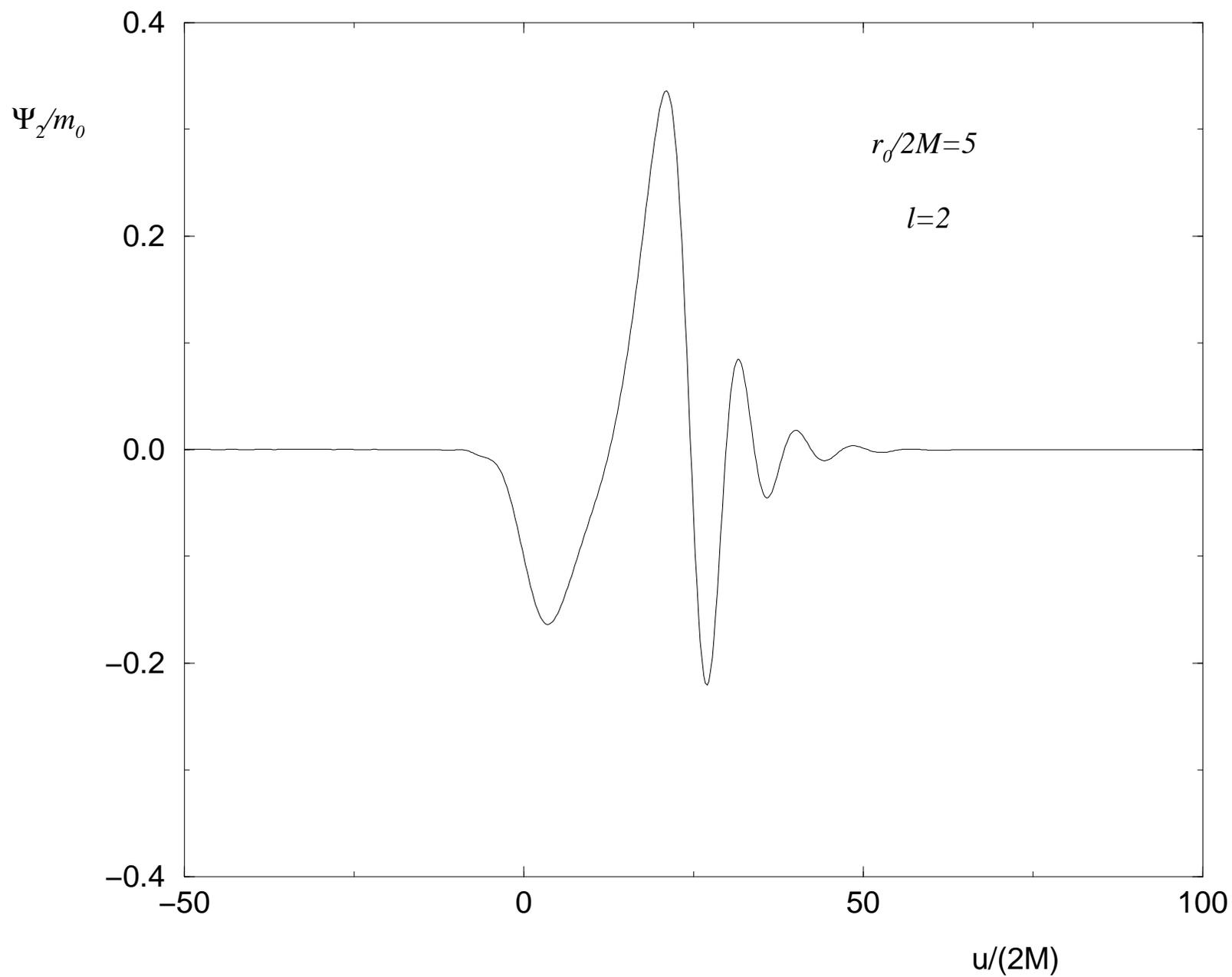
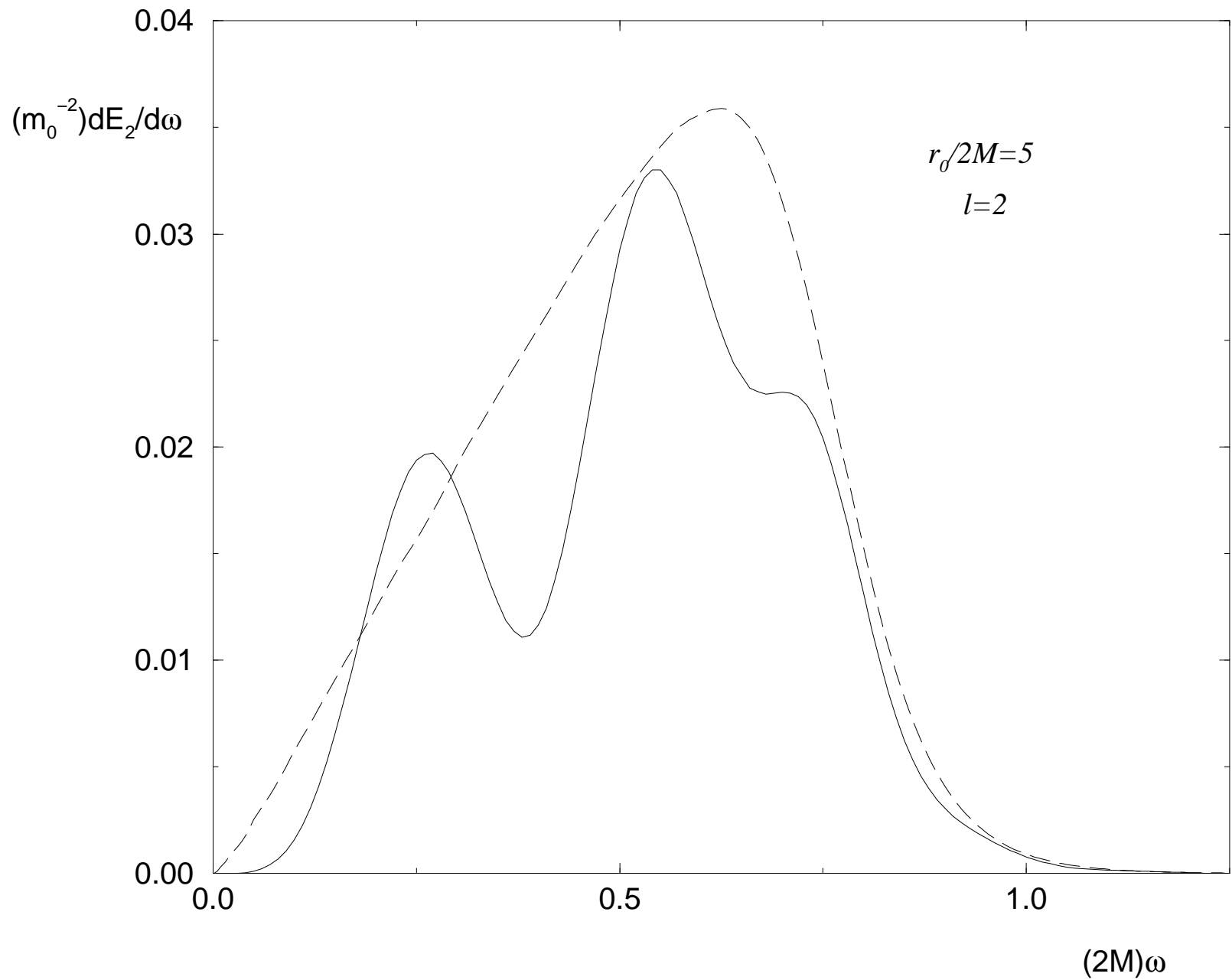
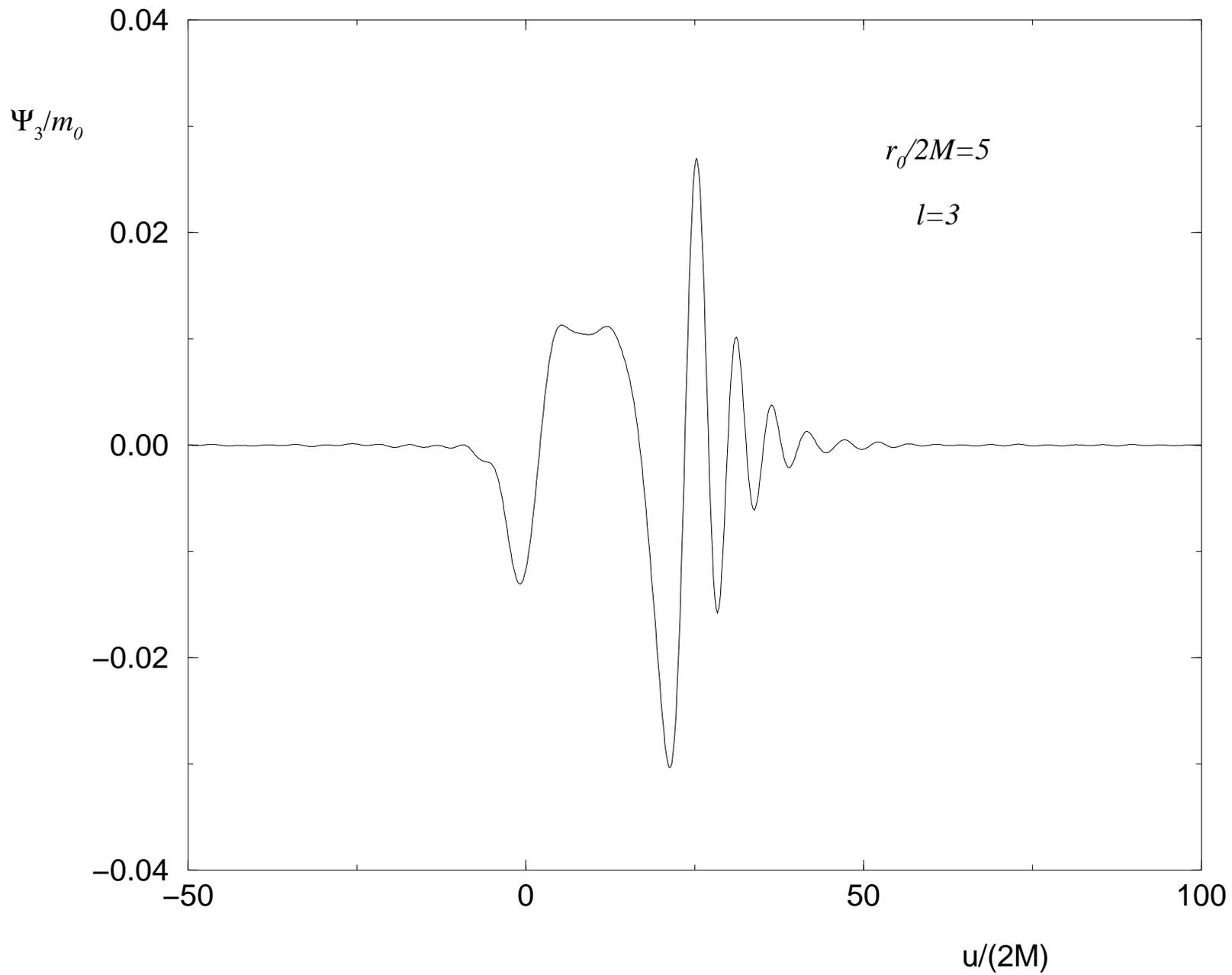
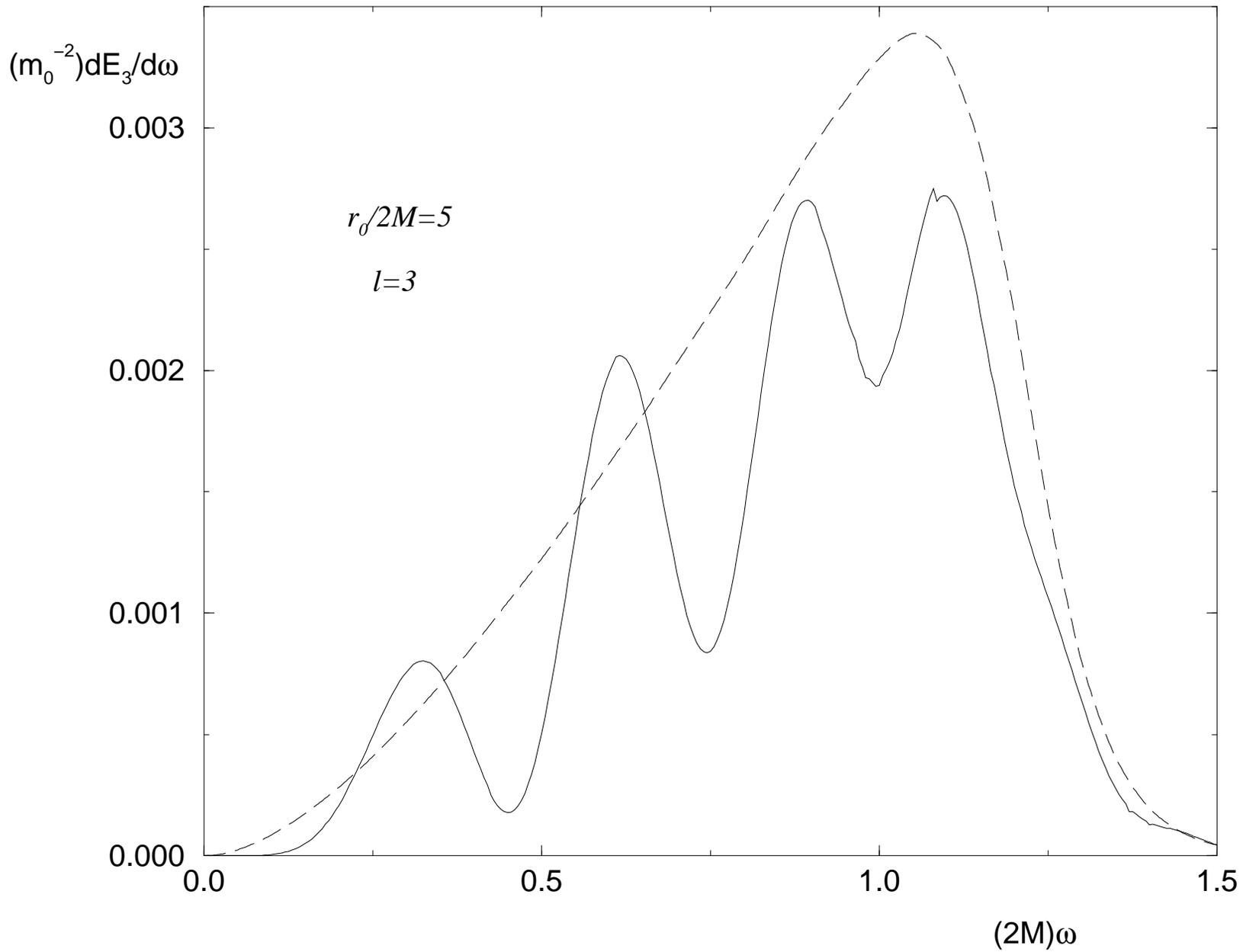
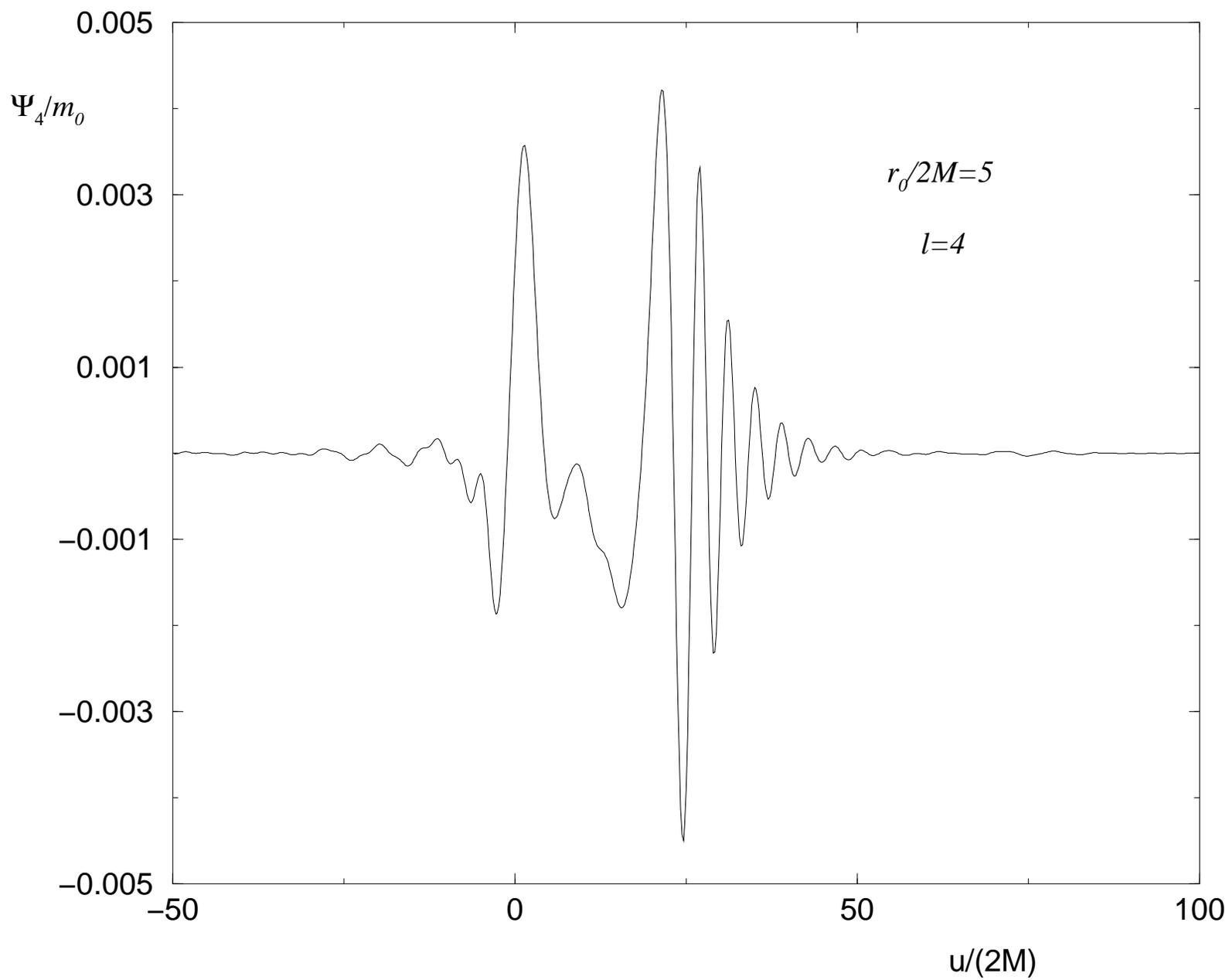
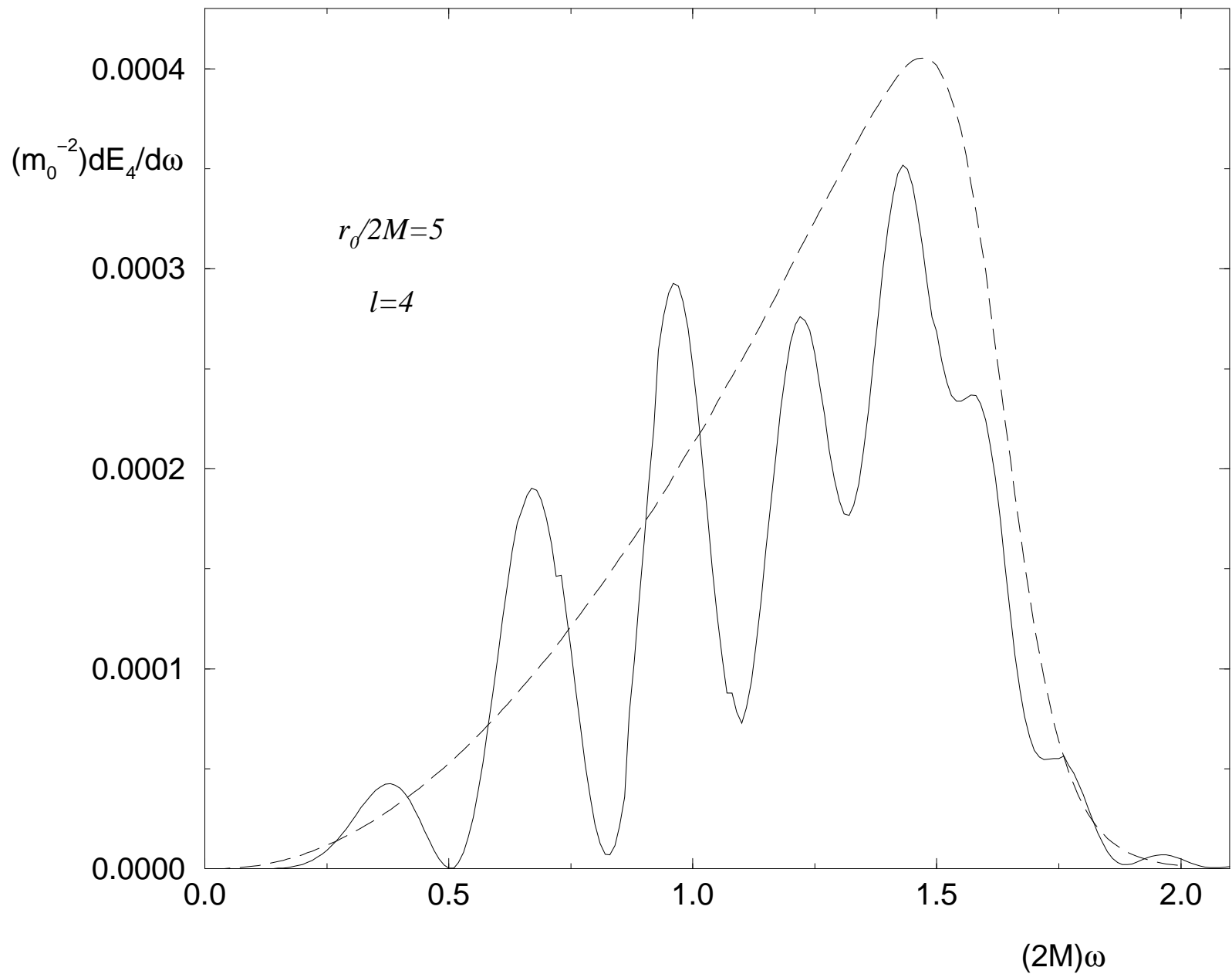
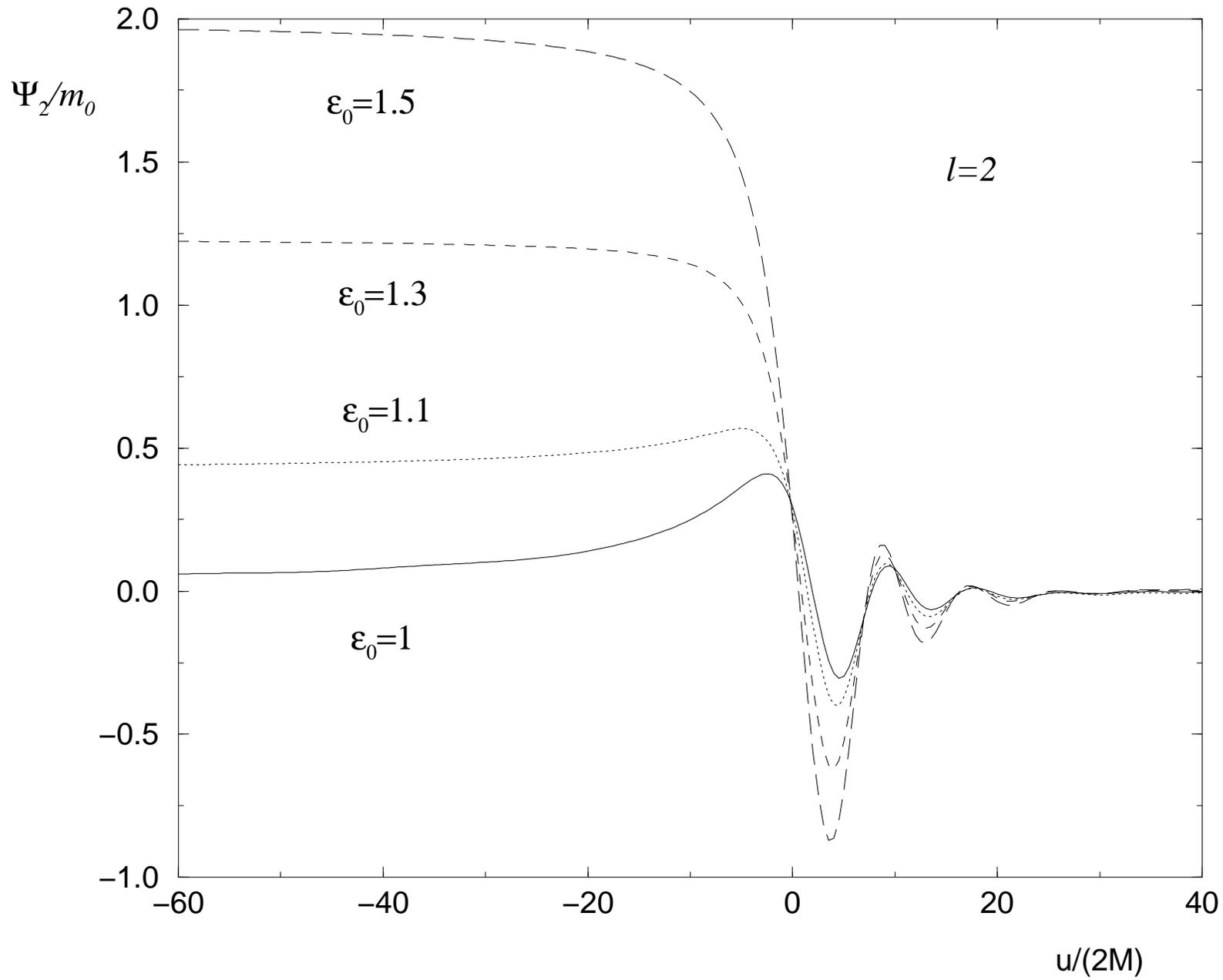
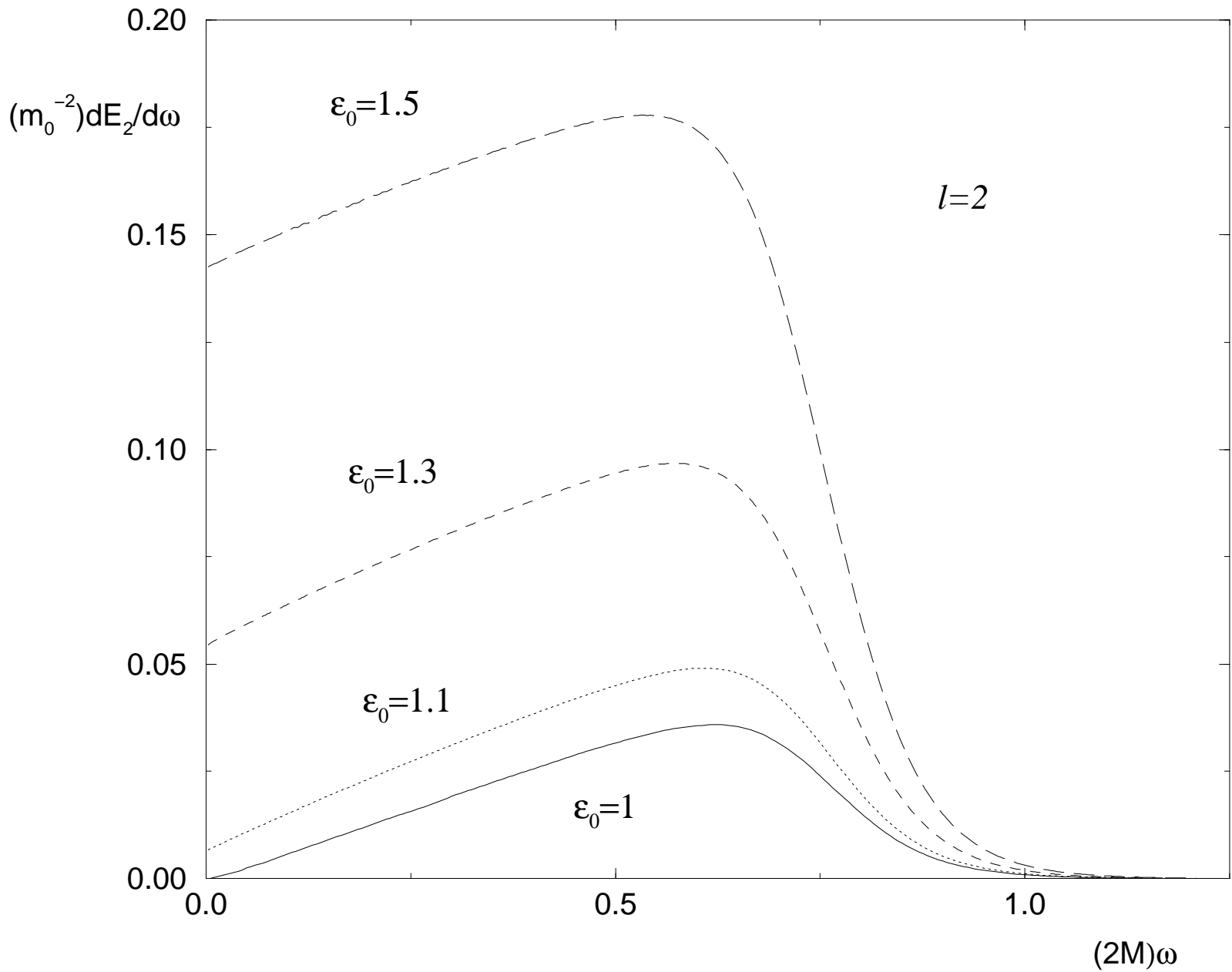
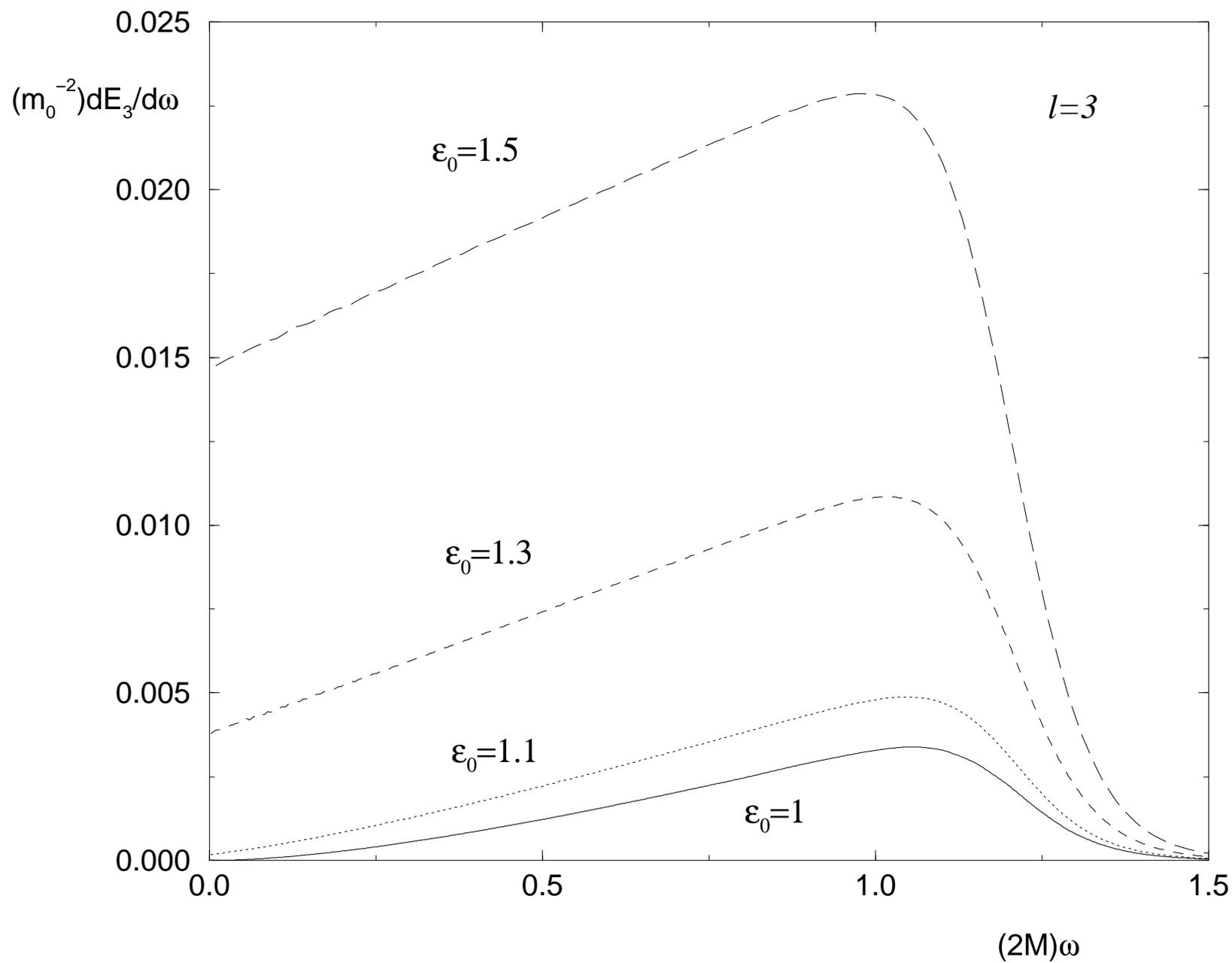
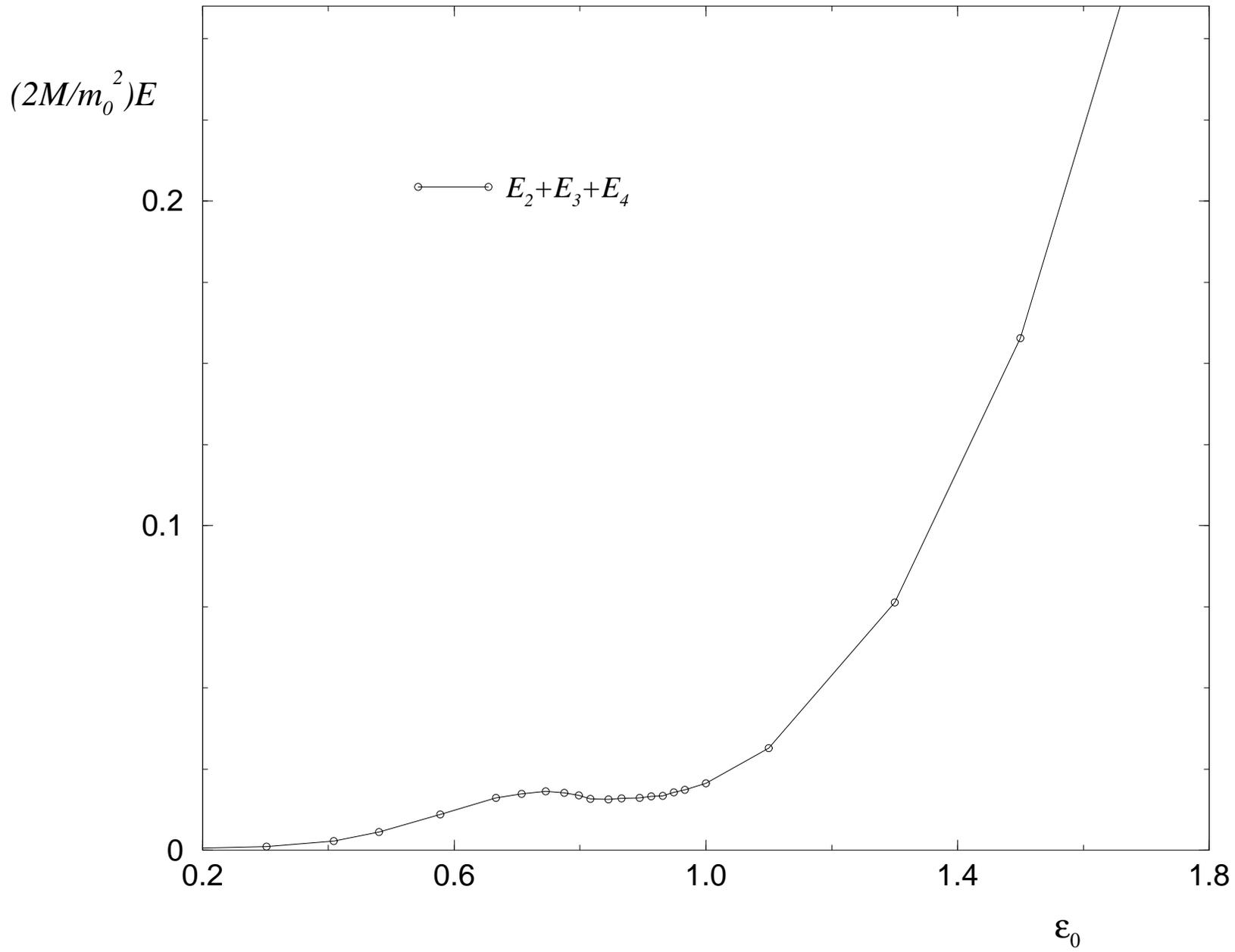
\end{document}